\documentclass[conference]{IEEEtran}
\usepackage{booktabs} 
\usepackage{makecell}
\usepackage{listings}
\usepackage{courier}
\usepackage{balance}
\usepackage[normalem]{ulem}
\usepackage[ruled,noline,noend,linesnumbered]{algorithm2e}
\SetAlFnt{\small}
\usepackage{multirow}
\usepackage{multicol}
\usepackage{graphicx}
\usepackage{subcaption}
\usepackage{enumitem}
\usepackage{amsmath}
\usepackage{amsthm}
\usepackage{amssymb}
\usepackage{comment}
\usepackage{xcolor}
\usepackage{url}
\usepackage{lineno}

\newcommand\skd[1]{{\color{black}#1}} 
\newcommand\sdn[1]{{\color{black}#1}} 
\newcommand\sdnj[1]{{\color{black}#1}} 

\definecolor{pinegreen}{rgb}{0.0, 0.47, 0.44}
\newcommand\sr[1]{{\color{black}#1}} 
\newcommand\srn[1]{{\color{black}#1}} 
\newcommand\srt[1]{{\color{black}#1}} 

\newcommand\LAI{FLAIRS}
\newcommand\ULI{TULSI}

\usepackage{caption}
\newtheorem{definition}{Definition}

\begin{document}
\title{Learned Adaptive Indexing with Updatable Learned Index}

\author{\IEEEauthorblockN{Suvam Kumar Das and Suprio Ray\\} 
\IEEEauthorblockA{\textit{Faculty of Computer Science, University of New Brunswick, Canada.} 
Email: \{suvam.das, sray\}@unb.ca}
}

\maketitle

\begin{abstract}
Indexes can significantly improve search performance in relational databases. However, if the query workload changes frequently or new data updates occur continuously, it may not be worthwhile to build a conventional index upfront for query processing. Applications involving large \sr{datasets may not benefit from building an index \textit{a priori}. In  exploratory data analysis, data scientists need to perform interactive analysis, which cannot afford the overhead of index construction.}
Building a conventional \sr{fine-grained index, like B$^+$-tree on the entire dataset can be very expensive since it may involve processing terabytes to petabytes of data. Even, the index itself may take large amount of storage}.  Moreover, with a traditional indexing, data ingestion tends to slow down in case of heavy updates as the index also needs to be updated. In these cases, adaptive indexing is a viable solution instead of traditional indexing. \textit{Adaptive indexing} is a technique in which an index gets built \textit{on the fly} as a byproduct of query processing.
%

In recent years, research in database indexing has taken on a new direction where machine learning models are employed for the purpose of indexing. These indexes, known as \textit{learned indexes}, can be more efficient compared to traditional indexes such as B+-tree in terms of memory footprints and query performance. \skd{However, a learned index has to be constructed upfront and requires training the model in advance, which becomes a challenge in dynamic situations when workload changes frequently.}
To the best of our knowledge, no learned indexes  exist yet for adaptive indexing. 
We propose a novel learned approach, called \LAI, for adaptive indexing.
It is built \textit{on the fly} as queries are submitted. \srn{We also introduce an updatable learned  index \ULI{} that is} utilized by \LAI{} for 
indexing data. 
\LAI{} can employ a query workload prediction technique that makes future workload projection based on past workload data. We have evaluated our learned adaptive indexing
approach against existing adaptive indexes  for various query workloads. Our results show that our approach performs better than others in most cases, offering significant improvement in query performance.
\end{abstract}


\section{Introduction}
\label{intro}

Indexes in a data system enhance data retrieval speed, improve performance, and ensure data consistency. 
For decades, 
B+-tree \cite{B+-tree} has been a dominant approach for indexing. 
%
Traditional 
indexes, such as the B+-tree are constructed prior to query execution and they require scanning the entire dataset once during the index construction. But in some scenarios,
the queries may be localized to a specific portion of a dataset, or the workload is dynamic and the dataset keeps on changing due to continuous  data updates. In these situations, it may not be  worthwhile to create an index on the entire dataset in a database table. Rather, indexes can be built \textit{on the fly} as a by product of query execution. Once a sufficient number of queries are executed, an index on the complete dataset is built as part of the query execution. This is known as \textit{adaptive indexing}. Some of the indexes built on this concept are \textit{database cracking} \cite{ Stochastic_database_cracking, Database_Cracking} and \textit{adaptive merging} \cite{Adaptive_Index_Relational_Keys_Adaptive_Merging, Adaptive_Merging}.

Adaptive indexing can 
be beneficial in big data applications where datasets are massive and workloads are unpredictable. It is expensive to build a traditional  \sr{fine-grained index, such as B$^+$-tree,} on the entire dataset, which \sr{may occupy terabytes to petabytes of storage}. Building an index on the entire dataset may not be worthwhile in 
many scenarios. \sr{First,  in exploratory analysis involving big datasets, where queries are not known in advance, data scientists typically would like to start querying the data without waiting for index construction. On the other hand, indexes are still required to entail interactive query response times over large datasets~\cite{maroulis2021adaptive, holanda2019progressive}. Second, in many analytical applications a database may store many years of data, but users usually query recent or `hot' data that is of interest to the user. Indexing the entire dataset that spans many years may require significant amount of storage.} Third,  when workloads are unpredictable, access patterns changes frequently, and user queries may focus on a specific region or columns in a dataset at a time.  Fourth, with  update-heavy workloads, data ingestion becomes slower since the index also needs to be updated simultaneously. \sr{Due to the costs of indexing large datasets, big data systems such as, Apache Spark~\cite{url:Spark}  and Apache Hive, usually do a full table scan or use a coarse-grained index~\cite{dymestifying_hive_index}, rather than supporting a B-tree style fine-grained index.} \sr{In this work we focus on adaptively indexing large datasets managed by database systems.} 


In recent years, researchers have developed indexes that utilize learned models.
Learned indexes leverage   techniques, such as Linear Regression \cite{FitingTree}, 
 Spline Interpolation \cite{RadixSpline} \srn{ and various machine learning models}. They are trained on a subset of the lookup keys to ``learn'' the Cumulative Distribution Function (CDF) of the keys. 
Then the model is used to predict the location given any key. 
Learned index approaches, such as the FITing tree \cite{FitingTree} and the PGM index \cite{PGM-Index} 
have been shown to be superior to B+-trees in terms of storage space requirement and search time. 
Learned indexes have also been incorporated in other domains \cite{FLIRT, LISA, LIMS}. 

However,  a learned index needs to be built upfront and it incurs additional cost associated with training the learned model. This becomes a challenge for dynamic situations where, the dataset is frequently updated or becomes stale and needs discarding whenever new batch of data arrives. Hence, it may not be worthwhile to create a learned index upfront under these scenarios, as the time spent on creating the index on dynamic data could rather be used to answer thousands of user queries. On the other hand, to the best of our knowledge, no work has been done to incorporate the advantages of learned approaches for incremental construction of adaptive indexes yet, which could result in enhanced query performance compared to the present adaptive indexes.

To address the limitations of existing approaches, we  propose a Framework for Learned Adaptive Indexing based on Radix Spline.  (\textbf{\LAI}). \LAI{} is a general purpose approach and can use any learned index for its construction. \srn{However, for improved performance of adaptive indexing with updates, we introduce The Updatable Learned Spline Index (\textbf{\ULI}). }
Following the concepts of adaptive indexing, \LAI{} is incrementally built 
as a by product of query execution. Each query tries to find all $x$ in the dataset, such that $x \in [l,h]$, where $l,h$ are the query boundaries. Unlike traditional learned indexes which creates an index on the entire dataset, in \LAI{} a learned index gets created on a partition of the input dataset depending on the query ranges. As an advantage, \LAI{} can 
maintain different  types of learned indexes in  different partitions, 
creating an ensemble of learned indexes. This can prove to be beneficial, for instance, in situations where a specific portion of the dataset is updatable, leaving the rest of the data untouched. In such a situation, \LAI{} can maintain an updatable learned index, such as 
\srn{our \ULI{}} for the updatable partition and static learned indexes such as RadixSpline \cite{RadixSpline} for the other partitions. 

Learned indexes require the data to be sorted, however, \LAI{} also works for queries on unsorted data. We partition and reorganize the input data based on the query boundaries such that all the $keys$ answering a query fall within a single partition. 
shown to enhance sorting time on large datasets. 

In  enterprise 
applications,
the queries in different workloads can follow  certain patterns. In these cases,
future queries may be predictable from the previous queries. We employ a statistical forecasting technique for query workload prediction, to predict the future queries using the 
past queries. 
Once the next batch of queries is forecasted, our index is updated based on them. 
This approach can be beneficial for workloads whose queries follow specific patterns and are  predictable based on the current queries. 

We have conducted extensive experimental evaluation to compare the performance of \LAI{} against other adaptive indexes for various query workloads on a synthetic dataset of 100M keys and a YCSB benchmark dataset. 
Our experiments show that \LAI{} outperforms the other approaches for most of the workloads. We also
demonstrate that with query workload prediction 
\LAI{} 
achieves performance improvement of {5.2$\times$--19.3$\times$ for some of the workloads, compared to the approach that do not use workload prediction.

To summarize, the key contributions of the paper are:
\begin{itemize}
    \item We propose \LAI{} that can integrate learned index with adaptive indexing.
    \item We introduce an updatable learned spline index \ULI. 
    \item We present a query workload prediction approach that can predict the next batch of queries from the past batch, and  update the index a priori for the predicted workload batch. 
    \item We conduct extensive experimental evaluation of \LAI{}  \srn{involving a number of different query workload scenarios.} 
\end{itemize}

The rest of the paper is organized as follow. In Section \ref{sec:related_work} we mention about the related work. Then in Section \ref{sec:updatablelearnedindex} we present \ULI{}, and in Section \ref{sec:learned_adaptive_index} we introduce \LAI{} and give a description of different query scenarios. Section \ref{sec:our_approach_and_algorithms} explains our 
algorithms.
Section~\ref{sec:query_prediction} mentions about query workload forecasting. 
We discuss  experimental evaluation in Section \ref{sec:exp_evaluation} and  conclude in Section \ref{sec:conclusion}. 
    
\section{Related Work}
\label{sec:related_work}
In this section, the related work is presented in detail. 

\subsection{Learned Access Methods}

\subsubsection{Learned Index}
RMI \cite{LearnedIndex} was the first learned index to be proposed, which is a hierarchical structure of neural networks. But this introduced space-time tradeoffs due to the complexity of the ML models adopted. To address this, FITing-Tree \cite{FitingTree} was introduced that uses a simple linear model at the leaf level of a B+-tree index. 
Another limitation of RMI is that it cannot be constructed over a single pass of the input data. To address this researchers proposed RadixSpline (RS) \cite{RadixSpline}, the first single-pass learned index with a \textit{constant} complexity to insert each element. RS utilizes a spline based interpolation method and it is not only efficient to build, but also competitive with state-of-the-art RMI models in size and lookup performance. 

\srn{The aforementioned learned indexes are built on static datasets and they do not handle updates efficiently. This limitation renders them unusable for dynamic, read-write workloads, common in many use cases. Recently, several updatable learned indexes were proposed, such as, ALEX~\cite{ALEX}, PGM~\cite{PGM-Index}, LIPP~\cite{LIPP}, (DILI)~\cite{DILI}, SALI~\cite{SALI} and RUSLI~\cite{RUSLI}.
}  

\subsection{Adaptive Indexes}
Adaptive indexing techniques, such as \textit{database cracking} \cite{ Stochastic_database_cracking, Database_Cracking} and \textit{adaptive merging} \cite{Adaptive_Index_Relational_Keys_Adaptive_Merging, Adaptive_Merging} allow for building an index incrementally in response to queries, rather than building it upfront. Database cracking progressively partitions and sorts a column by quicksort while answering range queries. 
While database cracking works as an incremental quick sort, adaptive merging functions as an incremental external merge sort on block-access devices such as disks, in addition to main memory. 
This merge logic executes as a side effect of query execution. Database cracking has a low initialization cost, but converges to a full index slowly, whereas adaptive merging has a relatively high initialization cost but converges rapidly. A hybrid combination of the two \cite{Adaptive_Indexing} that applies cracking on the initial partitions and sorting on the final one was proposed to achieve both lightweight initialization and quick convergence. These approaches blindly cracked the data space based on each query ranges.
Hence these approaches worked well with random queries, but faltered in scenarios when successive queries ask for consecutive items.

To address this, researchers later proposed \textit{Stochastic Database Cracking} \cite{Stochastic_database_cracking}, where apart from the query region, the data space is also cracked at arbitrary positions. 
Based on this idea, several approaches were proposed that use different strategies for data space cracking. For instance, DDC recursively halves until the resulting piece is sufficiently small, whereas in DDR, the data is cracked based on random pivots. Other variants include DD1C, DD1R and MDD1R.
AICC1R and AICS1R are variants of Crack-Crack
(AICC) and Crack-Sort (AICS) methods \cite{Adaptive_Indexing}. 


\section{\ULI: Our Updatable Learned Index}
\label{sec:updatablelearnedindex}

\subsection{Background on RadixSpline}
The core idea of adaptive indexing is to build indexes \textit{on the fly} during the query execution. To integrate the features of adaptive indexing with learned index, it is essential that not only the learned index  answers queries efficiently, but also gets built or updated quickly. If the sum of index building time, update time and query execution time is 
reasonably low, this could ensure 
high 
throughput.

RadixSpline (RS) offers 
low index-building time but it does not support updates.  In RS, a set of spline points is selected from the the input dataset by using the \textit{GreedySplineCorridor}~\cite{GreedySplineCorridor} technique based on the error bounds. A \textit{radix table} of size $2^k$ is used where each entry $i$ of this table points to the first spline point whose radix (considering the first $k$ bits) is $i$. Given a search \textit{key}, the radix of the first $k$ bits is calculated, and using the \textit{radix table} the two nearby enclosing spline points are determined. Then a binary search is conducted to find the nearest two spline points $k_1$, $k_2$, which enclose \textit{key}. Then a linear interpolation on $k_1$ and $k_2$ gives us the predicted position of \textit{key}, and finally the last-mile search is conducted to get its actual position. RS is a read-only index and to incorporate inserts, RUSLI~\cite{RUSLI} was later proposed, which expanded on the core idea of RS to handle inserts by maintaining an overflow array corresponding to each spline point.

With both RS and RUSLI, the number of radix bits need to be fixed before creating the models. A smaller error bound will lead to generation of larger number of spline points and hence a higher value of $k$ is better and vice-versa. If there are a lot of spline points having the same radix, then the smallest spline point will get pointed by the radix table, leaving out the rest. In contrast, a larger $k$ may lead to unused entries in the radix table.

\subsection{Our Approach}

\begin{figure}[t]
\begin{center}
\includegraphics[width=0.9\linewidth]{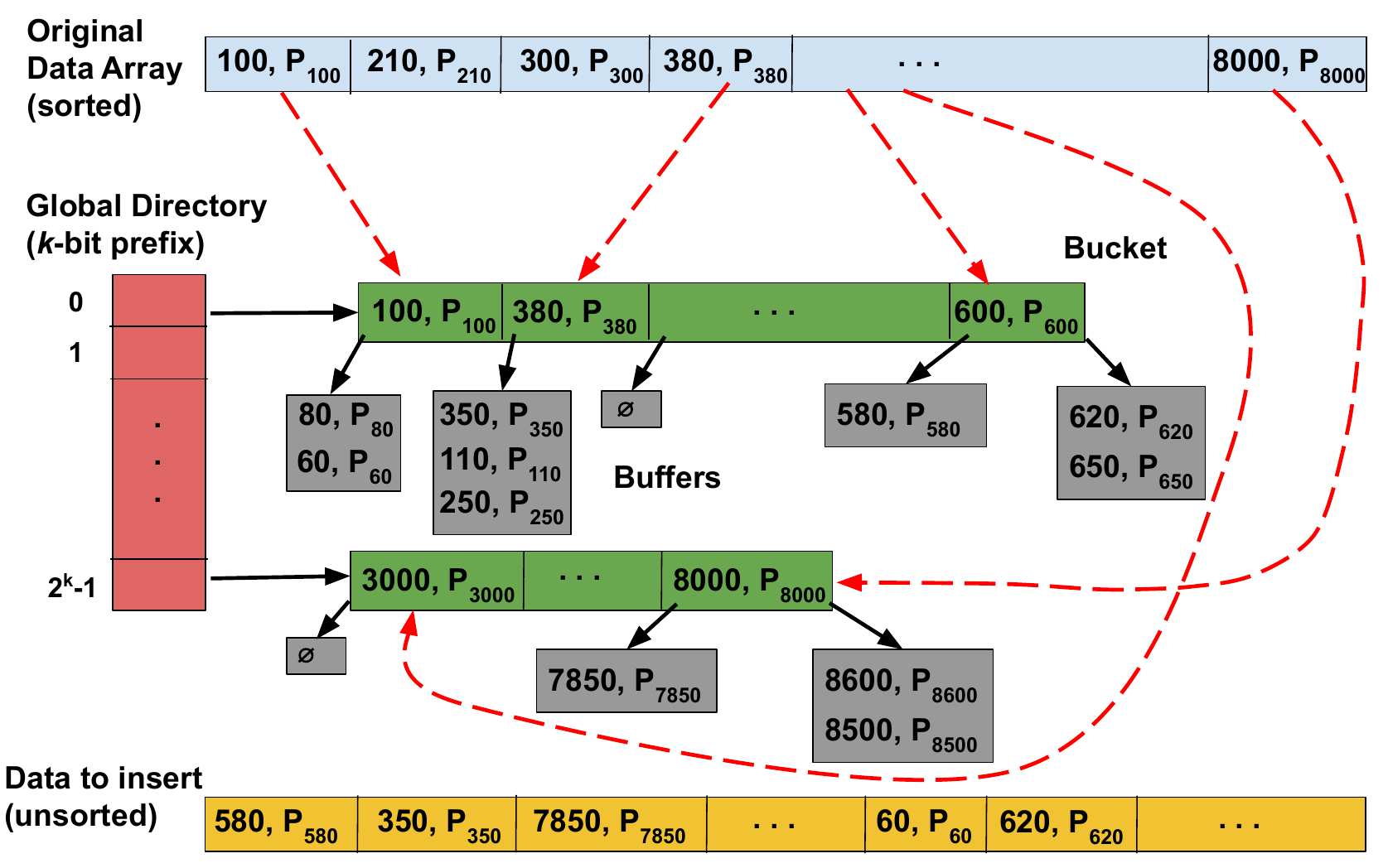}
\caption{\ULI{}  architecture}
\label{fig:TULSI_arch}
\end{center}
\end{figure}

To address these limitations, we introduce an updatable learned index \ULI, which is 
inspired by extendible hashing~\cite{FaginNPS79Extendible} and RS index. Figure \ref{fig:TULSI_arch} shows the architecture of \ULI. It uses an extendible hash table (EHT) to store the spline points. We initially start with a small $k$, which is equal to the \textit{global depth} of the \textit{directory} of extendible hashing. As more spline points are generated and added to the hash table, the \textit{global depth} keeps on increasing and we finally set the number of \textit{radix bits} as the global depth of the hash table, after all spline points are inserted. Each bucket of the hash table contains all spline points having the same $k$ first bits.

In contrast to the traditional approach for extendible hashing, where the \textit{least significant bits (LSB)} are considered to build the hash table,  we use the \textit{most significant bits (MSB)} for our purpose. If $MSB_{k}(key)$ denotes the radix using the first $k$ bits of $key$, then for any two keys $k_1$ and $k_2$, $k_1 < k_2 \implies MSB_{k}(k_{1}) \leq MSB_{k}(k_{2})$. Thus, if $k_1 \in B_1$ and $k_2 \in B_2$, where $B_1,B_2$ are the corresponding buckets, then all spline points in $B_1$ are smaller than all the spline points stored in $B_2$. Additionally, since the input data is sorted, the spline points are also generated in a sorted order and inserted into the bucket maintaining the same order. Since we use MSBs, the spline points gets added to the buckets pointed to by the lowest radix and gradually moving to higher radix bits. Hence all the spline points gets stored in the hash table in a globally sorted order.

Figure \ref{fig:TULSI_arch} shows an example to illustrate how \ULI{} works. Assume that we have an input data array of pairs $(key, P_{key})$, where $P_{key}$ denotes the position of $key$ in the array. This array is sorted by $key$. The sorted dataset $\{(100, P_{100}), (210, P_{210}), (300, P_{300}), ...\}$ is the input array (shown in blue). Using a similar approach as RS, we identify the spline points from the array, and insert them sequentially into the EHT. We extract the first $k$ bits and insert them into the bucket (shown in green) pointed to by the corresponding entry of the global directory (shown in red). For instance, assume the spline points $(100, P_{100}), (380, P_{380}),...,(600, P_{600})$ has the first $k$-bit prefix as 0 (for any arbitrary $k$). Hence all of them are stored in the bucket pointed to by the $0^{th}$ entry of the global directory (GD). Similarly, the spline points $(3000, P_{3000}),...,(8000, P_{8000})$ have their first $k$-bit prefix as $2^k-1$, hence they get stored in the bucket pointed to by the $(2^k-1)^{th}$ entry of GD. Whenever a spline point is tried to add to a full bucket, the entries in EHT may get redistributed following the principles of extendible hashing, and $k$ might get incremented by 1 if required. The red dashed arrows in the figure denote the spline points getting stored into different buckets based on their $k$-bit prefixes.

Next, let us assume that after building the EHT, a set of new data points are generated that needs to be inserted into \ULI{}.  In the figure, these new data points are $\{(350, P_{350}), (580, P_{580}), (7850, P_{7850}), ...\}$ and they  are shown inside the yellow array. To insert $(350, P_{350})$, we first calculate its first $k$-bit prefix, which happens to be 0. Then we do a binary search on the $key$ values stored in the bucket pointed to by the $0^{th}$ entry of GD to find the closest pair of spline points enclosing 350. In the example, the closest pair of spline points enclosing 350 is $(100, P_{100}), (380, P_{380})$. Then we insert $(350, P_{350})$ in the buffer pointed to by the spline point $(380, P_{380})$ (shown in gray), since each buffer associated with a spline point stores values less than the spline key. To minimize model re-training, only the last entry of a bucket stores an additional buffer (e.g. $(600, P_{600})$), which stores all elements greater than the last entry of this bucket and less than the first entry of the next bucket. Since model retraining is expensive, to avoid that as much as possible, this buffer is utilized to store keys when the buffer associated with the first entry of the next bucket is full. Empty buffers are denoted by $\varnothing$ in the figure.

\begin{algorithm}[t]
\caption{InsertNewKey$(val)$}
\label{algo:TULSI_insert}

$prefix := MSB_k(val)$ \;
$B_i = getBucket(prefix)$ \;
$(e_i,e_j) = getSplineEntries(B_i, val)$ \;
\If {$e_i$ and $e_j$ $\in B_i$} {
    \If{$e_j.b_l$ is not full} {
        $insert(e_j.b_l, val)$\;
    }
    \Else{
        $retrainModel()$\;
    }
}
\Else{
    \If{$e_j.b_l$ is not full} {
        $insert(e_j.b_l, val)$\;
    }
    \ElseIf {$e_i.b_h$ is not full} {
        $insert(e_i.b_h, val)$\;
    }
    \Else{
        $retrainModel()$\;
    }

}

\end{algorithm}

Each entry $e$ in a bucket is a pair $e=(s, b_l)$, except for the last entry, which is a triple of the form $e=(s, b_l, b_h)$, where $s$ is a spline point and $b_l, b_h$ are buffers respectively storing the values lower and higher  than $s$. The algorithm to insert a new key is shown in Algorithm \ref{algo:TULSI_insert}. To insert a new data item $val$, we first calculate $MSB_{k}(val)$ to find its radix (or prefix) \textbf{(Line 1)} and then fetch the corresponding bucket $B_i$ \textbf{(Line 2)}.
We perform a local binary search in $B_i$ to identify the pair of consecutive entries $(e_i, e_j)$, which contain spline points $(s_i, s_j)$ such that $s_i<val<s_j$ \textbf{(Line 3)}. Finally, we insert $val$ to the buffer $b_l$ of $e_j$, since $val < s_j$ \textbf{(Lines 4-6)}. If the buffer is full, the model needs to be retrained to accommodate $val$ \textbf{(Lines 7-8)}. 
In certain scenarios, $e_i$ and $e_j$ may belong to two different buckets. This situation arises when $val$ falls in between the last entry of a bucket ($e_i$ here) and the first entry of the next bucket ($e_j$ in this situation).  In such a case, we first try to insert $val$ to $b_l$ of $e_j$ \textbf{(Lines 10-11)}. 
If the buffer is full, we move to $e_i$, which is the last entry in the previous bucket and try to insert in its $b_h$ \textbf{(Lines 12-13)}. We need to retrain the index if this buffer is also full \textbf{(Lines 14-15)}. For retraining, we sort all elements in $b_l$ (or $b_h$) and merge it with the input sorted data array to create a new sorted array which further used to rebuild the index.

\begin{algorithm}[t]
\caption{SearchKey$(key)$}
\label{algo:TULSI_search}

$prefix := MSB_k(val)$ \;
$B_i = getBucket(prefix)$ \;
$(e_i,e_j) = getSplineEntries(B_i, val)$ \;
$s_i, s_j = getSplinePointsFromEntries(e_i,e_j)$\;
$posOrigArray = getPositionByRS(s_i, s_j, key)$\;
\If {$posOrigArray \neq -1$} {
    \Return $posOrigArray$\;
}
\If {$e_i$ and $e_j$ fall in same bucket} {
    $posInsArray = LinearSearch(e_j.b_l, key)$\;
}
\Else{
    $posInsArray = LinearSearch(e_i.b_h, key)$\;
    \If {$posInsArray = -1$} {
        $posInsArray = LinearSearch(e_j.b_l, key)$\;
    }
}
\Return $posInsArray$\;
\end{algorithm}

The algorithm to search a $key$ is quite similar to data insertion, which is depicted in Algorithm \ref{algo:TULSI_search}. Here again, we first calculate $MSB_{k}(key)$ to get its radix. Then we fetch the corresponding bucket $B_i$, to identify the closest enclosing spline point pairs $(s_i,s_j)$ from bucket entries $(e_i,e_j)  $\textbf{(Lines 1-4)}. The search $key$ could either belong to the original data array (sorted) or the inserted data array (unsorted). After identifying $(s_i,s_j)$, we proceed to search the original data region, in the same way that RS does \textbf{(Line 5)}. If the $key$ is found in the original data region, we return its position \textbf{(Lines 6-7)}. Else, we issue a search operation in the corresponding buffers of the spline points \textbf{(Lines 8-14)}. If $e_i$ and $e_j$ belong to the same bucket, then we perform a linear search on $e_j.b_l$ \textbf{(Lines 8-9)}. Otherwise $e_i$ is the last entry of a bucket, and $e_j$ is the first entry of the next bucket. In that case, a $key$ could either belong to $e_i.b_h$ or $e_j.b_l$. We first perform a linear search on $e_i.b_h$. If the $key$ is not found there, then we perform a linear search on $e_j.b_l$ \textbf{(Lines 10-13)}. Finally, we return the position of $key$ in the inserted data array \textbf{(Line 14)}. After identifying $(s_i,s_j)$, since both the search techniques are independent of each other \textbf{(Lines 5-7 and Lines 8-14)}, they can be executed in parallel using multi-threading to enhance performance.  For example, in Fig. \ref{fig:TULSI_arch}, to search 300, the first $k$-bits are used to identify the bucket containing the enclosing spline points. In this case it is the bucket stored at the $0^{th}$ entry of GD, and the enclosing spline points are $(100, P_{100}), (380, P_{380})$. 
We use these points to find $P_{300}$ from the original data array. To search 350, we follow the same approach and find that it does not exist in the data array. In that case we perform a linear search on the buffer of spline point $(380, P_{380})$ and return  $P_{350}$ in the inserted data array.

To search a range of values falling in between two keys $k_1$  and $k_2$, the following approach is taken. For each of the two keys, we get a pair of entries, $(e_1,e_2)$ for $k_1$ and $(e_3, e_4)$ for $k_2$ having spline point pairs $(s_1,s_2)$ and $(s_3, s_4)$ respectively, where $s_1 < k_1 < s_2$ and $s_3 < k_2 < s_4$. We need to find all $x$ such that $k_1 \leq x \leq k_2$, where $x$ either belongs to the data array or the inserted array. Using
$(e_1,e_2)$ and $(e_3,e_4)$, we identify all keys $k'$ residing in the original data array such that $k_1 \leq k' \leq k_2$ and append them to the final result set. To identify all inserted keys within the range, we then traverse through all entries of the EHT, which fall between $e_2$ and $e_4$. We append to the final result, all the buffer entries between $e_2$ and $e_3$ (not including $e_2$ but including $e_3$). Finally, we search for all entries greater than $k_1$ in $e_2.b_l$ and less than $k_2$ in $e_4.b_l$ and append it to the final result set. The corresponding algorithm for this
is not shown due to space constraint.

\begin{figure}[t]
\begin{center}
\includegraphics[width=0.7\linewidth]{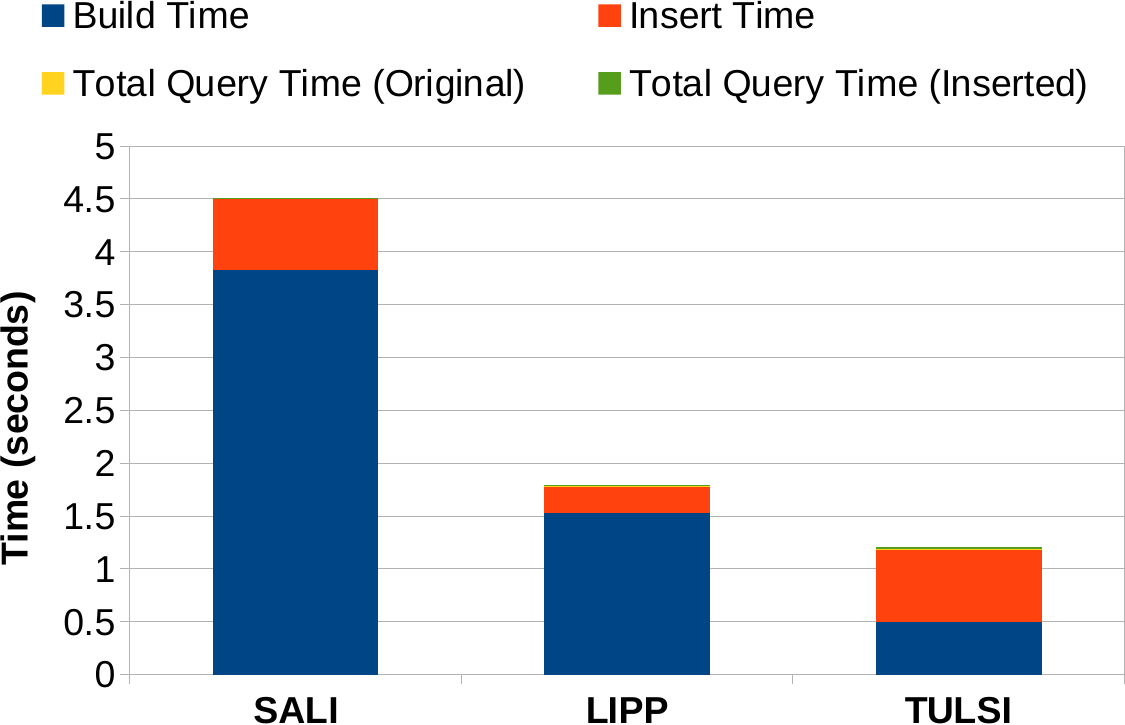}
\caption{Evaluation of \ULI{}}
\label{fig:eval_tulsi}
\end{center}
\end{figure}

\subsection{Evaluation of \ULI{}}

\noindent We implemented \ULI{} in C++ and performed an evaluation on a micro-benchmark against two recently proposed updatable learned indexes, SALI
~\cite{SALI} and LIPP
~\cite{LIPP}. We could not compare \ULI{} with 
RUSLI~\cite{RUSLI} since its source code was not publicly available. 

We generated a random integer dataset (called, `Original') of 20M keys and sorted them. We  built all the three learned index model on the dataset. Then we generated 2M random integer keys (called, `Inserted') and inserted them into the 
indexes.
Finally we generated 20K point search queries in which 10K queries belonged to the original dataset and the remaining 10K to the newly inserted dataset. We repeated the entire process three times and measured in each iteration the \textbf{index build time} on the Original dataset of 20M keys, \textbf{insertion time} of all the 2M keys and \textbf{total query time on Original} and the \textbf{total query time on Inserted}. Finally, we individually computed the average of all the measurements, which is shown as a stacked graph in Fig. \ref{fig:eval_tulsi}. To accommodate inserts, each of the spline points stored in the buckets of the extendible hash table had buffers of size 100. It was not possible to tune the paremeters for LIPP and SALI, and so 
we used their default settings. By default, SALI maintains an initial memory pool of size 100. Our experiments reveal that
\ULI{} performs worse than SALI and LIPP with point search queries only. Whereas, \ULI{} performs significantly better than both SALI and LIPP \sdn{on the combined total time of the three criteria}, namely, the total time to build a learned index on the initial data, time to update and the query execution time.
For the index building on the Original dataset, SALI and LIPP take 3.8 and 1.5 secs respectively, whereas \ULI{} takes 0.5 secs. For insertion, LIPP takes the least time, which is 0.25 secs, whereas SALI and \ULI{} almost take the same time of 0.68 secs. For answering point search queries, the performance remained similar for all the three indexes, irrespective of the query points residing in the Original or the Inserted data. In this case, SALI and LIPP performed better compared to \ULI{}, where SALI and LIPP take around 5 millisecs whereas \ULI{} takes 12 millisecs.
However, the query execution time for a point search query is significantly less that those of insertion and update times, due to which they are not clearly visible in Fig. \ref{fig:eval_tulsi}. The total time taken is around 1.2 secs for \ULI{}, whereas LIPP and SALI take  1.8 and 4.5 secs respectively.  \LAI{} can use any updatable learned index. However,
since the learned index model needs to be built, updated and searched \textit{on the fly} during query execution, 
an updatable index with minimal total \textit{end-to-end} time (i.e. sum of building, insertion and query time) is ideal. Hence, we utilize \ULI{} as a learned index model for \LAI{}.

\section{\LAI: Our Approach For Learned Adaptive Index}
\label{sec:learned_adaptive_index}

\begin{figure}[t]
\begin{center}
\includegraphics[width=0.9\linewidth]{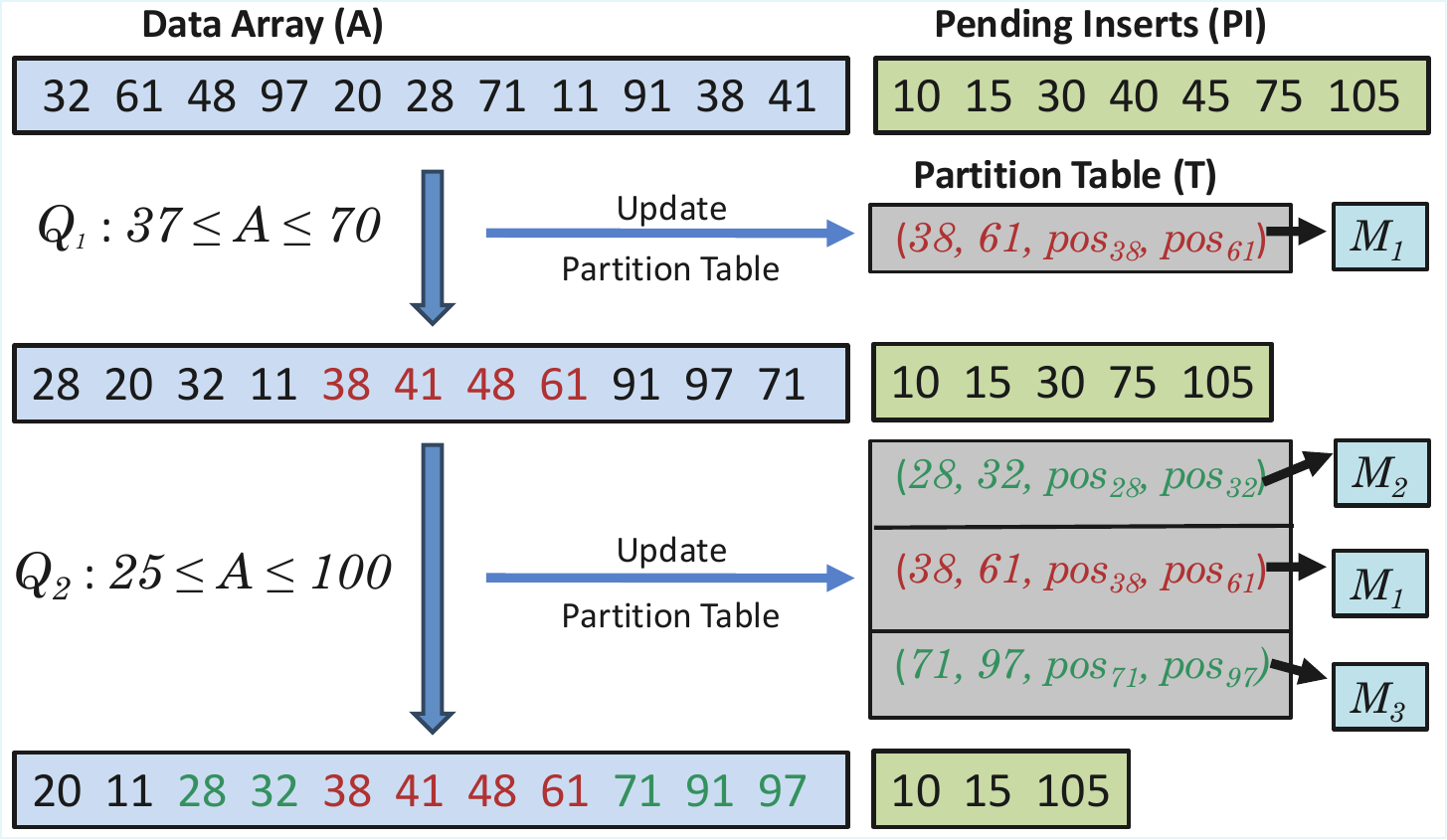}
\caption{An illustration of adaptive indexing with \LAI}
\label{fig:LAI_basic_idea}
\end{center}
\end{figure}

Inspired  by the idea of learned indexes and 
the advantages in terms of storage consumption and query execution over the traditional indexes, we propose to integrate 
learned indexes with adaptive indexing.
We call this approach \LAI{}, which 
can efficiently process \textit{Range Queries}. A Range Query is defined as follows.

\begin{definition}[Range Query]
Given an array $A$ (sorted or unsorted) and an interval $[l,h]$, find all $x \in A$, such that $l \leq x \leq h$.
\end{definition}

The basic concepts behind \LAI{} are illustrated in Fig. \ref{fig:LAI_basic_idea}. 
Let $A$ be an unsorted array and $PI$ be the sorted set of data pending to be inserted. A \textit{Range Query $Q_1$} is submitted, where $Q_1:37 \leq A \leq 70$. The idea is to \textit{crack A} based on the query endpoints i.e. 37 and 70, such that three disjoint partitions $P_1, P_2$ and $P_3$ are created where $P_1$ contains all values $< 37$, $P_2$ contains all values within the range $[37, 70]$ (marked in red), and finally $P_3$ contains the rest of the values greater than $70$. As can be seen in the figure, $P_2$ answers $Q_1$. We then do a binary search on $PI$, and remove all entries satisfying $Q_1$ from $PI$. Let this be $I_1 = \{40, 45\}$. The result of this query is $P_2 \cup I_1$, which is returned to the user. This partition $P_2$ is then sorted using a standard sorting algorithm, and a learned index $M_1$ is created on this partition. $M_1$ is then updated with $I_1$. Since the lowest and highest value in $P_2$ are 38 and 61 respectively, a tuple of the form $(38,61,pos_{38}, pos_{61}, M_1)$ is stored in a 1-D Partition Table $T$, where $pos_{38}$ and $pos_{61}$ are the positions of 38 and 61 respectively in $A$. When another query $Q_2:25 \leq A \leq 100$ is submitted, we scan $T$ to identify that 25 lies in $P_1$ whereas 100 belongs to $P_3$. It is easy to identify the start and end indexes of the partitions $P_1, P_3$ from T, since an entry in $T$ stores both the key as well as its position in $A$. Next, we crack $P_1$ based on 25 to find all elements $\geq 25$ and sort them (shown in green) to create a learned model $M_2$.  Then we check $PI$ to remove all entries falling within $[25, 38)$, 38 being the start point of the next partition. Let that be $I_2=\{30\}$. We further update $M_2$ based on $I_2$. We see that 28 and 32 are the two endpoints of this sorted partition. We create a tuple of the form $(28,32,pos_{28},pos_{32},M_2)$ and insert it into T. Similarly we crack $P_3$ based on 100 and find all keys $\leq 100$ and then sort them (shown in green) to create another learned model $M_3$. We again check $PI$ to remove all entries within $(61,100]$, 61 being the end point of the previous partition. Let this be $I_3 = {75}$. We further update $M_3$ with $I_3$. We again find that the end points of this sorted partition are 71 and 97. So, we create another tuple $(71,97,pos_{71},pos_{97},M_3)$ and insert it to $T$. Finally to answer $Q_2$ all elements in $A[pos_{28}:pos_{97}] \cup I_2 \cup I_3$ are returned. In order to improve search time, the entries in $T$ are inserted in such a way that they remain sorted by the boundary interval $[l,h]$. Thus the tuple $(28,32,pos_{28},pos_{32},M_2)$ appears before $(38,61,pos_{38}, pos_{61}, M_1)$, whereas $(71,97,pos_{71},pos_{97},M_3)$ appears after it. As and when more queries arrive, $A$ gets further refined and new entries are inserted in $T$, until enough queries are submitted so that an index gets created on the entire array $A$. 
Subsequently submitted queries can be directly  answered by using $T$ only. \sdn{This creates several learned models spanned across the entire data space, which can possibly be merged to create a few \srt{models} or a single model. However we do not take this approach, since models need to be merged to create a new model during query execution, 
} 
\srt{incurring additional overhead.} When one or both of the end points of a query already fall(s) within a sorted partition, its corresponding learned model is consulted to obtain the position(s) of the 
key(s). However, as and when new records arrive, $PI$ gets updated, and depending on the query boundaries data may get removed from $PI$ and inserted into a learned model. In the event of model retraining, the data is inserted into the model are merged with the original sorted partition for which the learned model was initially built. In that situation, the entries in $T$ gets updated.
Next, we discuss our approach in detail.


\subsection{Handling different query  scenarios}
\label{subsec:Diff_query_scenarios}
\label{sec:handlingquery}
\begin{figure}[t]
  \begin{subfigure}[b]{0.47\textwidth}
	   \centering
	   \includegraphics[width=\linewidth]{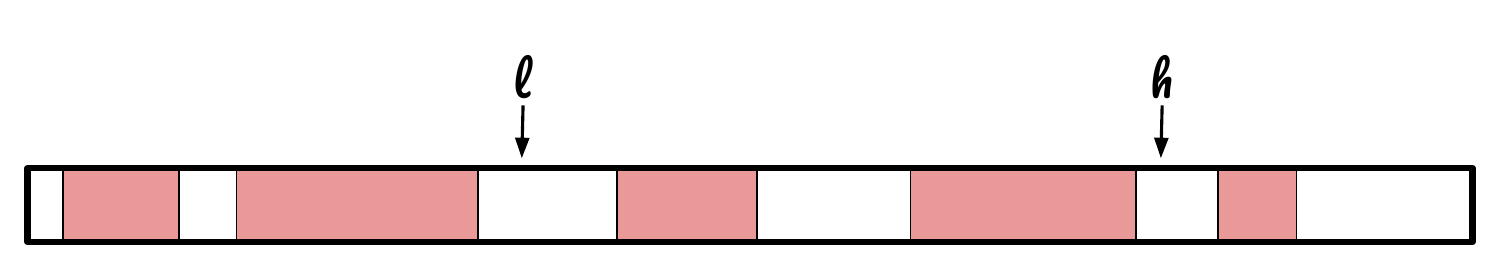}
	    \vspace{-5pt}
	    \caption{Case 1(i): $l$ and $h$ overlap existing partition(s)}
	    \label{subfig:overlap_query}
	\end{subfigure}    
  \begin{subfigure}[b]{0.47\textwidth}
	   \centering
	   \includegraphics[width=\linewidth]{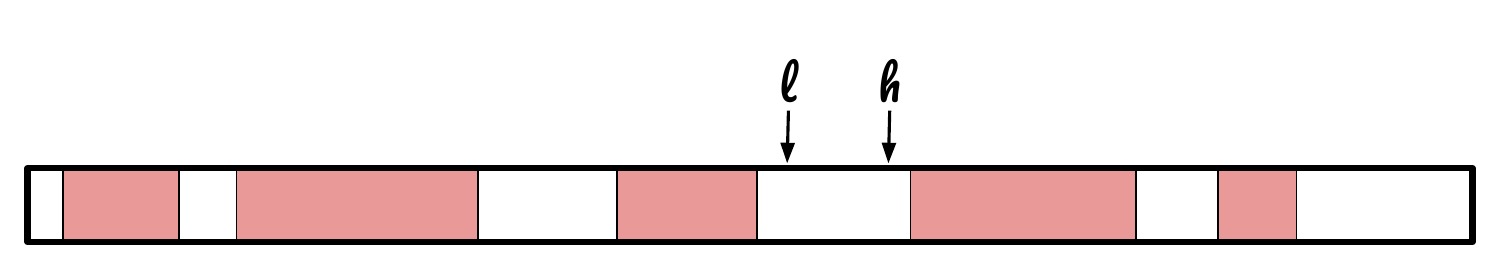}
	    \vspace{-5pt}
	    \caption{Case 1(ii): $l$ and $h$ fall within a single unsorted partition}
	    \label{subfig:build_index}
	\end{subfigure}    
  \begin{subfigure}[b]{0.47\textwidth}
	   \centering
	   \includegraphics[width=\linewidth]{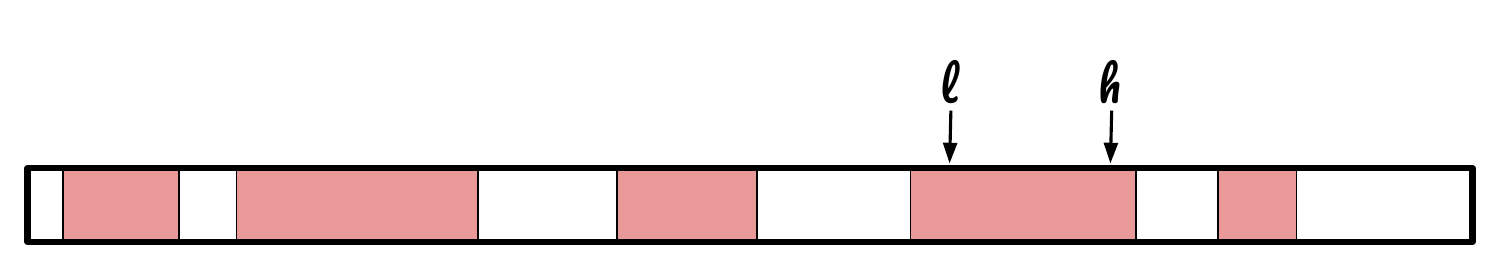}
	    \vspace{-5pt}
	    \caption{Case 2: $l$ and $h$ fall within a single sorted partition}
	    \label{subfig:same_partition}
	\end{subfigure}    
  \begin{subfigure}[b]{0.47\textwidth}
	   \centering
	   \includegraphics[width=\linewidth]{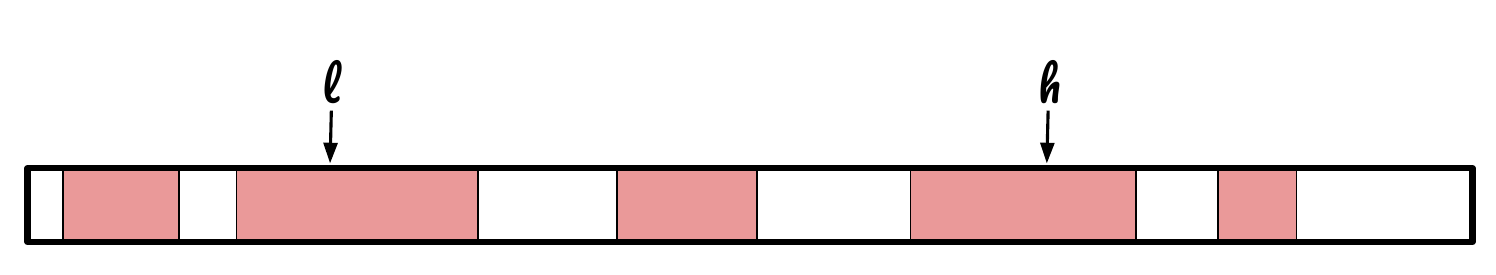}
	    \vspace{-5pt}
	    \caption{Case 3: $l$ and $h$ fall in two different sorted partitions}
	    \label{subfig:different_partition}
	\end{subfigure}    
  \begin{subfigure}[b]{0.47\textwidth}
	   \centering
	   \includegraphics[width=\linewidth]{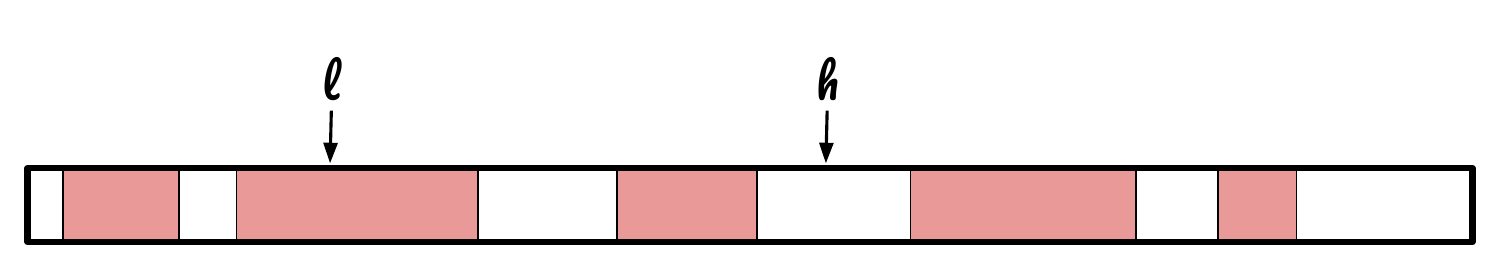}
	    \vspace{-5pt}
	    \caption{Case 4: Only $l$ falls within a sorted partition}
	    \label{subfig:crack_high}
	\end{subfigure}    
 \hspace{15pt}
  \begin{subfigure}[b]{0.47\textwidth}
	   \centering
	   \includegraphics[width=\linewidth]{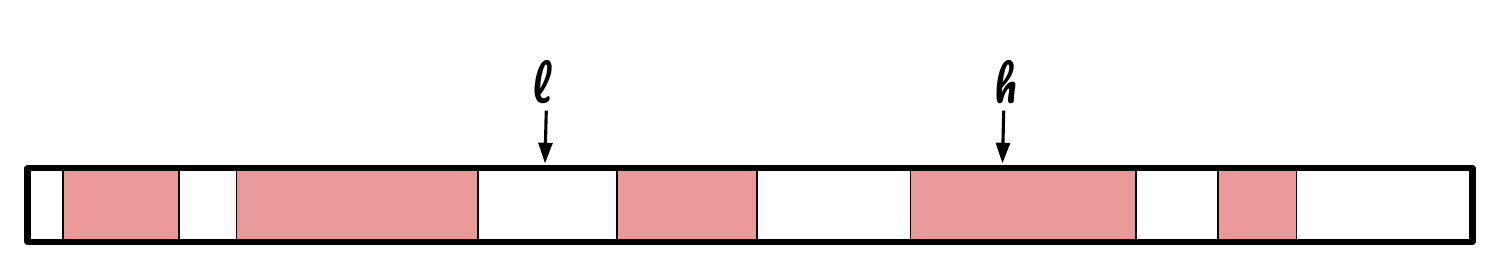}
	    \vspace{-5pt}
	    \caption{Case 5: Only $h$ falls within a sorted partition}
	    \label{subfig:crack_low}
	\end{subfigure}    
\caption{Different query scenarios}
\label{fig:different_query_scenarios}
\end{figure}

For a query $Q$ to find all $x \in A$ , such that $l \leq x \leq h$ is invoked, one of the following cases 
holds, as shown in Fig. \ref{fig:different_query_scenarios}. Please note, we assume that pending inserts array (PI) is available but is not shown here, due to space limitations.
In the figure, the black outer rectangle represents the entire data range, whereas the red solid rectangles represent those partitions that have been sorted by previous queries, and learned indexes is/are constructed for those partitions. The white spaces represent unsorted partitions, which have not been touched by any query yet. Note that, before any query arrives, $A$ is considered to be a single partition:\\
\textbf{Case 1}: Both $l$ and $h$ reside within an unsorted partition. This can be divided into two sub-scenarios:
\begin{enumerate}[label=\roman*.]
    \item Sorted partition(s) already exists between $l$ and $h$. This happens when both the end points of $Q$ cover or overlap existing sorted partition(s), as shown in Fig. \ref{subfig:overlap_query}.

    \item No sorted partition exists between $l$ and $h$, i.e. both endpoints belong to a single unsorted partition, as shown in Fig. \ref{subfig:build_index}. This situation always arises when a query is issued for the first time on $A$. Or there might be existing sorted partitions before and/or after this query range.
\end{enumerate}

\noindent\textbf{Case 2}: Both $l$ and $h$ belong to the same sorted partition, as shown in Fig. \ref{subfig:same_partition}. \\
\textbf{Case 3}: Both $l$ and $h$ belong to different sorted partitions, as shown in Fig. \ref{subfig:different_partition}. \\
\textbf{Case 4}: Only $l$ belongs to an already sorted partition, but $h$ does not, as shown in Fig. \ref{subfig:crack_high}. \\
\textbf{Case 5}: $l$ does not belong to an already sorted partition, but only $h$ does, as shown in Fig. \ref{subfig:crack_low}.\\
\sr{Next, we discuss} each scenario in detail. For 
this discussion, we  assume that each scenario 
is initialized as follows:

\begin{figure}[t]
\vspace{-8pt}
\centering
\includegraphics[width=0.7\linewidth]{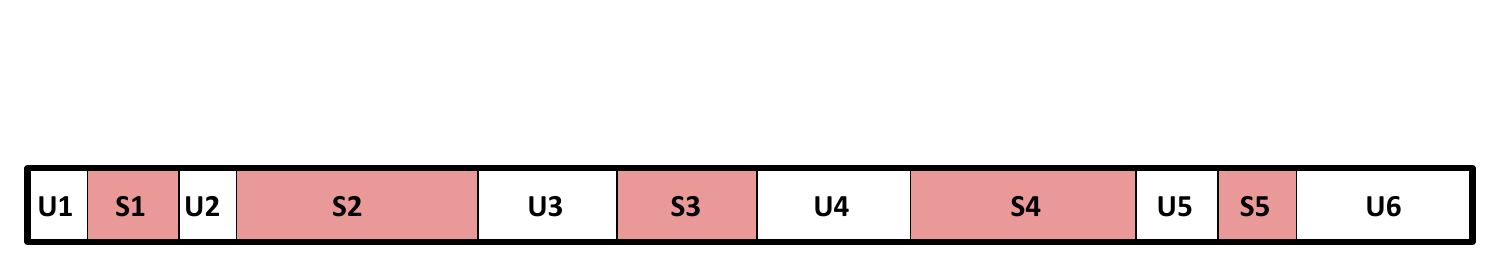}
\caption{Example scenario}
\label{fig:example_situation}
\vspace{-4pt}
\end{figure}

\begin{itemize}
    \item The array $A$ has been divided into disjoint sorted and unsorted partitions as a result of queries submitted prior to $Q$ as shown in Fig. \ref{fig:example_situation}. There are 6 unsorted partitions $U_1-U_6$ (shown in white) and 5 sorted partitions $S_1-S_5$ (shown in red). Although in our example, we always have an unsorted partition between two sorted partitions, in real it is possible to have two contiguous sorted partitions without an unsorted partition in between, as we  
    discuss later in this section.

    \item For each sorted partition $S_i$, a tuple of the form $T_i \equiv$ ($l_i$, $h_i$, $pos_{l_i}$, $pos_{h_i}$, $M_i$) is stored in the partition table $T$, where $l_i, h_i$ are the corresponding low and high values of $S_i$, $pos_{l_i}, pos_{h_i}$ are the locations of $l_i, h_i$ in  $A$, and $M_i$ is the corresponding learned index 
    (we use \ULI{} here).
    The entries of $T$ are sorted by the the interval $[l, h]$ i.e. $[l_i, h_i] < [l_j, h_j]$ implies $T_i$ appears before $T_j$. It is evident, that for all $x_i, x_j \in S_i, S_j$, if $S_i$ appears before $S_j$ in $A$, then $x_i < x_j$. This is because, before creating each $S_i$, it was first \textit{cracked} based on $l_i$, followed by $h_i$ ensuring that all values $x_i \in S_i, l_i \leq x_i \leq h_i$ always holds. 

    \item If $S_i$ appears before $U_j$ in $A$, this implies that for all $x_j \in U_j, \mbox{ } x_j > h_i$. Conversely, if $S_i$ appears after $U_j$, it implies that for all $x_j \in U_j, \mbox{ } x_j < l_i$. Thus if $S_i, U_j, S_k$ appears in order, then $h_i < x_j < l_k$, for all $x_j$. 
    Furthermore, if $U_i$ appears before $U_j$, then $x_i < x_j$ for all $x_i,x_j \in U_i, U_j$.
\end{itemize}

An important consideration is \textit{how to identify, whether one or both the endpoint(s) of a query reside(s) in a sorted or an unsorted partition.} This can be determined by scanning the \textit{Partition Table T}. Since $T$ stores the intervals of the sorted partition in a sorted manner, performing a binary search over the stored interval can give us the stored partition in which the endpoint(s) reside. If no such partitions exist, then it must belong to an unsorted partition.

\subsection{Handling updates}
\label{sec:handlingupdates}

In order to handle updates in \LAI, we keep track of all inserts in a \textit{Pending Insert (PI)} buffer. The entries are kept in sorted order to reduce query time. Once a new range query \textit{Q} is submitted, we check \textit{PI} to search whether there are keys falling within that range. All such keys are reported to the user and then are finally inserted either to an existing or a newly created learned model depending on the query scenario as described in Section \ref{subsec:Diff_query_scenarios}. We use \ULI{}  as an updatable learned index for this purpose. Initially, as queries are processed, and partitions from the array gets sorted, \ULI{} is used to create a learned model over this partition. Then $PI$ is checked for any keys falling within the query range and are  inserted into the \ULI{} model corresponding to the current partition. When another query arrives whose range coincides with an existing partition, we retrieve the relevant \ULI{} based learned model and use it to answer range searches. We also insert new data to it (if there are any in $PI$ falling within the partition boundary). \sdn{Currently, deletion is not supported, which 
}
\srt{is a future work.}

\section{Algorithms}
\label{sec:our_approach_and_algorithms}

As new queries are submitted, we consult the \textit{Partition Table T}, to identify the scenario \textbf{(Cases 1 -- 5)} 
that is applicable for this query. 
Then, the algorithm 
corresponding to that case is invoked. Next we discuss each case in details along with the algorithms. 
We assume that the following helper methods exist:

\begin{itemize}
    \item $searchForSortedPartitions(T,l,h)$: This method searches the partition table $T$ to find those partitions in which $l$ (or $h$) may lie, and return the positions ($T_l,T_h$) of those partitions in T. If $l$ (or $h$) does not lie inside any partition, then $T_l$ (or $T_h$) is returned as -1. 
    \item $isOverlapQuery(T,l,h)$: This method scans $T$ and checks whether the query boundaries $l$ and $h$ span across one or multiple sorted partitions or not. If yes, then it returns \textit{True}, otherwise \textit{False}.
    \item $searchForGap(T,k)$: Given a key $k$, scan $T$ and return the boundaries ($p_1,p_2$) of the gap in which $k$ lies.
    \item $crack(p_1,p_2,k)$: Crack i.e. rearrange the sub-array A[$p_1:p_2$] on $k$, such that all values $< k$ are on the left, while values $\geq k$ are on the right of $k$.
    \item $getLearnedIndex(T, pos)$: Given a position $pos$, this method extracts the stored tuple $(l,h,pos_l,pos_h,M)$ at $T[pos]$ and then returns learned model $M$. 
    \item $getBoundaries(T, pos)$: Given a position $pos$, this method extracts the stored tuple $(l,h,pos_l,pos_h,M)$ at $T[pos]$ and then returns the boundary $(pos_l,pos_h)$.
    \item $getAllGaps(T, p_1, p_2)$: Given two indices $p_1$ and $p_2$ of the global array $A$, return the list of $gaps$, such that each $gap$ $U$ is denoted in the form $U \equiv (U_1,U_2)$, where $(U_1,U_2)$ are the boundaries of $U$. 
    \item $find(M,k)$: Given a key $k$, this method finds the position of $k$ in the global array A, using the learned model $M$.
    \item $findInsertedDataFromModel(M, l, h)$: Given $l$ and $h$, this method retrieves all elements falling within that range from model $M$ which were later inserted to it. 
    \item $findInPendingInserts(l,h)$: This method retrieves all elements from $PI$ falling within $[l, h]$.
    \item $updateModel(M, L)$: Insert all elements $l \in L$ into 
    $M$.
    \item $getHighValue(T,pos)$:Given a position $pos$, this method extracts the stored tuple $(l,h,pos_l,pos_h,M)$ at $T[pos]$ and then returns $A[pos_h]$.
    \item $getHighValuePos(T,pos)$:Given a position $pos$, this method extracts the stored tuple $(l,h,pos_l,pos_h,M)$ at $T[pos]$ and then returns $pos_h$.
    \item $getLowValue(T,pos)$:Given a position $pos$, this method extracts the stored tuple $(l,h,pos_l,pos_h,M)$ at $T[pos]$ and then returns $A[pos_l]$.
    \item $insertDataBetweenBoundaries(pos_l, pos_h)$: Given $pos_l$ and $pos_h$ of $T$, this method finds all data entries in $PI$ that fall between two sorted partitions in $A$ and for each of those partitions, it inserts values to their corresponding learned model $M$, and returns the list of data retrieved from $PI$.

\end{itemize}

\begin{algorithm}[t]
\caption{query$(l, h)$}
\label{algo:query}
  \KwIn{A query $Q$ to find all $x$ from a global array $A$, such that $l \leq x \leq h$
  }
  \KwOut{$Result$ satisfying the query boundary}
$(T_l,T_h) := searchForSortedPartitions(T,l,h)$\;
\If{$T_l$ = -1 and $T_h$ = -1} {
    \If{isOverlapQuery(T,l,h) = True} {
        $Result := executeOverlapQuery(l,h)$ //Case 1(i)\;
    }
    \Else {
        $Result := buildIndex(l,h)$ //Case 1(ii)\;
    }
}

\ElseIf{$T_l$ = $T_h$} {
    $Result := getResultsFromSameBound(l,h,T_l)$ //Case 2\;
}

\ElseIf{$T_l \geq 1$ and $T_h \geq 1$} {
    $Result := getResultsFromDiffBound(l,h,T_l,T_h)$ //Case 3\;
}

\ElseIf{$T_l \geq 1$ and $T_h = -1$} {
    $Result := crackForHighValue(l,h,T_l)$ //Case 4\;
}

\Else {
    $Result := crackForLowValue(l,h,T_h)$ //Case 5\;
}

\Return $finalResult$ \;
\end{algorithm}

The driver algorithm \textit{query}, as shown in Algorithm \ref{algo:query}, is executed every time a query is submitted. Based on the query boundaries $(l,h)$ and the current state of $T$, any one of the cases 1--5 is satisfied and the corresponding algorithm gets invoked. From the query boundaries we try to identify the entry in $T$, which may contain $l$ (and $h$) and get the position of entries $(T_l,T_h)$ \textbf{(Line 1)}. If such a position is not found for both the boundaries (i.e. $T_l=T_h=-1$) \textbf{(Lines 2--6)}, we try to identify and act accordingly if the query belongs to \textbf{Case 1(i)} \textbf{(Lines 3--4)} or \textbf{Case 1(ii)} \textbf{(Lines 5--6)}. If $T_l=T_h$, then \textbf{Case 2} is invoked \textbf{(Lines 7--8)}. The query belongs to \textbf{Case 3} when both $T_l,T_h \geq 1$ \textbf{(Lines 9--10)}. If $T_l \geq 1$, but $T_h=-1$, then the scenario belongs to \textbf{Case 4} \textbf{(Lines 11--12)}. If none of the conditions are satisfied, then the situation falls under \textbf{Case 5} \textbf{(Lines 13--14)}. Finally, the query results are returned \textbf{(Line 15)}. As a by-product of this query the \textit{Partition Table T} gets further updated.

\subsection{Case 1: Both $l$ and $h$ resides in an unsorted partition}
As mentioned earlier, this scenario can arise in two situations, which are described 
next.

\subsubsection{\textbf{Case 1(i) : Sorted partition(s) already exists between $l$ and $h$}}

\begin{figure}[t]
\vspace{-12pt}
\begin{center}
\includegraphics[width=0.7\linewidth]{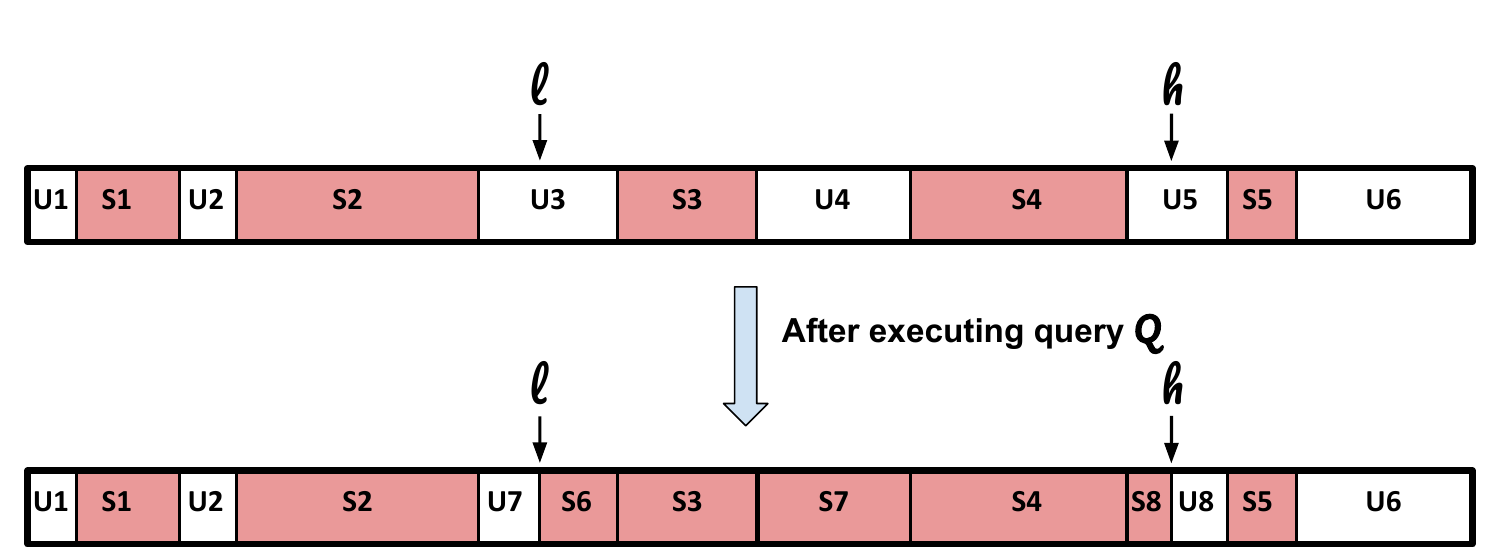}
\caption{Sorted partition(s) already exists between $l$ and $h$}
\label{fig:overlap_query_2}
\end{center}
\end{figure}

\begin{algorithm}[t]
\caption{executeOverlapQuery$(l, h)$}
\label{algo:executeOverlapQuery}
  \KwIn{A query $Q$ to find all $x$ from a global array $A$, such that $l \leq x \leq h$
  }
  \KwOut{$finalResult$ satisfying the query boundary}
$(p_l,q_l) := searchForGap(T,l)$ //Get the indices for gap $U_l$\;
$(p_h,q_h) := searchForGap(T,h)$ //Get indices for gap $U_h$\;
$res1 := crackForLowValue(l,A[q_l],\_)$ \;
$res2 := crackForHighValue(A[p_h],h,\_)$\;
$res3 := buildIndexForAllGaps(q_l,p_h)$ \;
$(pos_l, pos_h) := searchForSortedPartitions(T,l,h)$\;
$res4 := insertDataBetweenBoundaries(pos_l, pos_h)$ \;
$finalRes := res1 \cup res2 \cup res3 \cup res4$\;
\Return $finalRes$ \;
\end{algorithm}

\begin{algorithm}[t]
\caption{buildIndexForAllGaps$(pos_1, pos_2)$}
\label{algo:buildIndexForAllGaps}
  \KwIn{$pos_1, pos_2$ are two indices
  }
  \KwOut{Build index for all gaps between $pos_1$ and $pos_2$ and update the partition table $T$}

$gapList := getAllGaps(T, pos_1, pos_2)$\;
\For {$(p_1,p_2) \in gapList$} {
    $lowValue = min(A[p_1:p_2])$ //Get min value in gap\;
    $highValue = max(A[p_1:p_2])$ //Get max value in gap\;
    $buildIndex(lowValue,highValue)$ //Build index\;
}
\Return \;
\end{algorithm}

This scenario is depicted in Fig. \ref{fig:overlap_query_2} and Algorithm \ref{algo:executeOverlapQuery} shows the steps in details. Let us assume that $l$ and $h$ reside in $U_3$ and $U_5$ respectively, which have $S_3,U_4$ and $S_4$ in between. First, the partition table $T$ is scanned (as stated above) to find the unsorted partitions in which $l$ and $h$ reside. 
They are denoted by $U_l$ and $U_h$ (in this case it is $U_3$ and $U_5$). Let $p_l, q_l$ be the lower and upper indexes for $U_l$, and $p_h,q_h$ for $U_h$ \textbf{(Lines 1--2)}. Next, we \textit{crack} the partition $U_l$, i.e the sub-array $A[p_l:q_l$] on $l$ . Performing this operation results in $U_l$ to be shuffled so that all values $\geq l$ appear to the right of $l$. Here we sort the sub-array $A[l_{pos}:q_l]$ and create a learned index (model) $M_l$. We store the low and high values of the sorted sub-array along with $M_l$ in $T$ \textbf{(Line 3)}. Similarly, $U_h$ is cracked based on $h$ ensuring that all values $\leq h$ fall on the left of $h$ \textbf{(Line 4)}. \sdn{The steps for line 4 ($crackForHighValue$) is described in Algorithm  \ref{algo:crackForHighValue}}. Again, we sort $A[p_h:h_{pos}]$, and create a learned index $M_h$ and add the corresponding entries to $T$. Let the positions of $l$ and $h$ after \textit{cracking} $U_l$ and $U_h$ be $l_{pos}, h_{pos}$ respectively. By this approach, the original unsorted $U_l$ gets refined into an unsorted partition on the left of $l_{pos}$ and a sorted partition on the right of $l_{pos}$ (e.g., $U_3$ gets refined into $U_7$ and $S_6$). The effect is similar for $U_h$ too 
. 
Then, we check for any whole unsorted partitions falling within $q_l$ and $p_h$ by searching $T$, which we denote as \textit{gaps}. The \textit{gaps} can  be identified from $T$, by taking into account of the fact that if two sorted partitions are contiguous, then the higher index of the first partition should be one less than the lower index of the second partition. If that is not the case, then there are $gaps$ in between the two sorted partitions. 
Once all such gaps within $q_l$ and $p_h$ are identified (in our example, there is only one \textit{gap} which is $U_4$), they are sorted and a learned model is built over them and the necessary informations are added to $T$ (in our example, $U_4$ is converted to $S_7$) (\textbf{Line 5}, details shown later in Algorithm \ref{algo:buildIndexForAllGaps}). Then we insert all data from $PI$ falling across partition boundaries \textbf{(Lines 6--7)}. Finally we combine all intermediate results and return them \textbf{(Line 8)}.

Algorithm \ref{algo:buildIndexForAllGaps} shows the steps for building a learned model over all the gaps between the boundaries $pos_1$ and $pos_2$. \skd{We get the list of all unsorted partition boundaries between $pos_1,pos_2$ \textbf{(Line 1)}. We iterate through all the unsorted boundaries \textbf{(Lines 2--5)} and do the following for each of them. We identify the minimum \textbf{(Line 3)} and maximum \textbf{(Line 4)} for a partition and then build an index for this range and update $T$ accordingly  \textbf{(Line 5)}.} Algorithm \ref{algo:buildIndexForAllGaps} is used in the other cases as well, as we will see next.

\subsubsection{\textbf{Case 1(ii) : Both $l$ and $h$ belong to the same unsorted partition}}

\begin{figure}[t]
\vspace{-12pt}
\begin{center}
\includegraphics[width=0.7\linewidth]{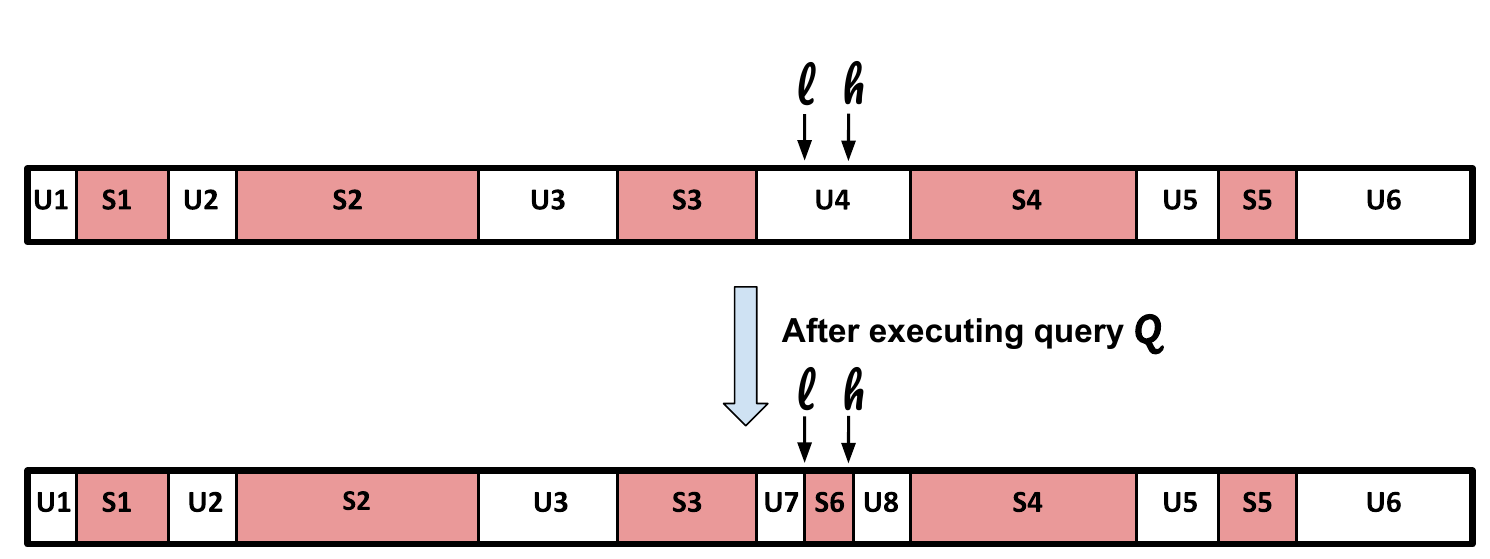}
\caption{Both $l$ and $h$ belong to the same unsorted partition}
\label{fig:build_index_2}
\end{center}
\end{figure}


\begin{algorithm}[t]
\caption{buildIndex$(l, h)$}
\label{algo:buildIndex}
  \KwIn{$l,h$ are query boundaries. Build a learned index for the portion $P$ of the global array $A$ such that $l \leq x \leq h$, where $x \in P$
  }
  \KwOut{The pair ($l_{pos}, h_{pos}$), the two indices of $P$}
$(p_l,q_l) := searchForGap(T,l)$ //Get indices for the gap\;
$l_{pos} := crack(p_l,q_l,l)$ //Crack the gap on $l$\;
$h_{pos} := crack(l_{pos},q_l,h)$ //Crack the gap on $h$\;
$sort(l_{pos},h_{pos})$ //Sort the sub-array $A[l_{pos}:h_{pos}]$\;
$M := buildLearnedIndex(l_{pos},h_{pos})$\;
$resfromPI := findInPendingInserts(l,h)$\;
$updateModel(M, resfromPI)$ \;
$entry := (l,h,l_{pos},h_{pos},M)$\;
$insert(T,entry)$ //Update T sorted by [$l_{pos},h_{pos}$]\;
$finalResult := A[l_{pos}:h_{pos}] \cup resfromPI$\;
\Return $finalResult$ \;
\end{algorithm}

This is always the case when the first query is submitted, but this can occur in other scenarios too, as we see from Fig. \ref{fig:build_index_2}. Algorithm \ref{algo:buildIndex} shows this scenario. Let  both endpoints belong to an unsorted partition $U_i$ ( $U_4$ in our case). In such a situation, $U_i$ is broken into a sorted partition and possibly into multiple unsorted partitions by \textit{cracking} based on $l$ and $h$ \textbf{(Lines 1--3)}, and then sorting the sub-array \textbf{(Line 4)}, as shown in the previous scenario. Finally a learned index model is built over this new sorted partition \textbf{(Line 5)}. $PI$ is checked for values falling within the query range and $M$ is updated \textbf{(Lines 6--7)}. An entry is created for this partition and is added to $T$ \textbf{(Lines 8--9)}. Finally, all entries satisfying the query from $A$ and $PI$ are combined together and returned \textbf{(Lines 10--11)}. In our example, $U_4$ is broken into $U_7, S_6$ and $U_8$, and the necessary information for $S_6$ is added to $T$.

\subsection{Case 2: Both $l$ and $h$ belong to the same sorted partition}


\begin{algorithm}[t]
\caption{getResultsFromSameBound$(l, h, pos)$}
\label{algo:getResultsFromSameBound}
  \KwIn{$l,h$ are query boundaries. $pos$ is the index of partition in $T$ in which the query boundaries lie.
  }
  \KwOut{$finalResult$ satisfying the query boundary}
$M := getLearnedIndex(T, pos)$ //Get the learned index from T[pos]\;
$l_{pos} := find(M,l)$ //Get the position of $l$ from M\;
$h_{pos} := find(M,h)$ //Get the position of $h$ from M\;
$resFromModel := findInsertedDataFromModel(M, l, h)$\;
$resfromPI := findInPendingInserts(l,h)$\;
$updateModel(M, resFromPI)$\;
$finalResult := A[l_{pos}:h_{pos}] \cup resFromModel \cup resfromPI$\;
\Return $finalResult$\;
\end{algorithm}

This 
scenario is depicted in Fig. \ref{subfig:same_partition} and this is a
simple case  to handle as there is no need to reorganize the data. Algorithm \ref{algo:getResultsFromSameBound} shows the steps of this scenario. After issuing the query, we scan $T$ to find that both the query endpoints belong to a sorted partition $S_i$ (say). We select the corresponding learned index $M$ \textbf{(Line 1)} and use that to find the locations of $l$ and $h$ \textbf{(Lines 2--3)}. Next, we retrieve all data inserted into $M$ falling within $[l,h]$ \textbf{(Line 4)} and all data in \textit{PI} satisfying the query boundary \textbf{(Line 5)}. Then we update $M$ \textbf{(Line 6)} and then combine all the intermediate results to return the $finalResult$ \textbf{(Lines 7--8)}. 

\subsection{Case 3: Both $l$ and $h$ belong to different sorted partitions}

\begin{figure}[t]
\vspace{-12pt}
\begin{center}
\includegraphics[width=0.7\linewidth]{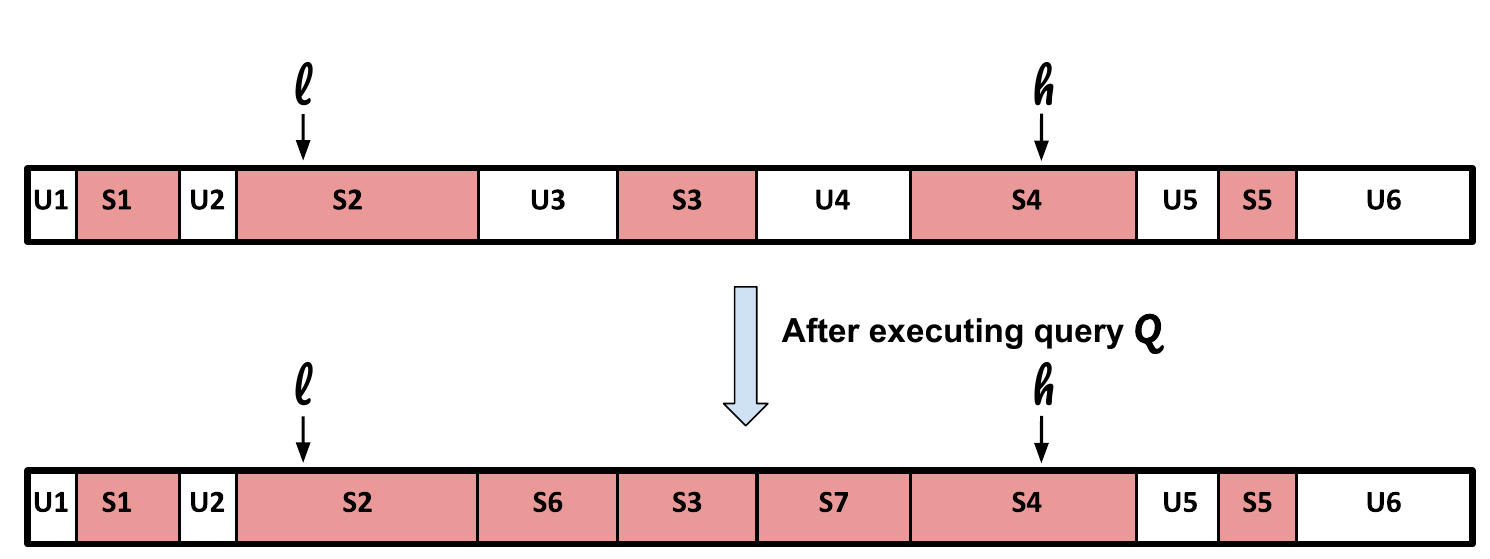}
\caption{Both $l$ and $h$ belong to different sorted partitions}
\label{fig:different_partition_2}
\end{center}
\end{figure}


\begin{algorithm}[t]
\caption{getResultsFromDiffBound$(l, h, pos_l, pos_h)$}
\label{algo:getResultsFromDifferentBound}
  \KwIn{$l,h$ are query boundaries. $pos_l, pos_h$ is the indices of partition in $T$ in which $l,h$ respectively lie.
  }
  \KwOut{$finalResult$ satisfying the query boundary}

$tempHigh := getHighValue(T, pos_l)$\; 
$res1 := getResultsFromSameBound(l,tempHigh,pos_l)$\;
$tempLow := getLowValue(T, pos_h)$\; 
$res2 := getResultsFromSameBound(tempLow,h,pos_h)$\;
$(p_l,q_l) := getBoundaries(T, pos_l)$ //Get boundaries with $l$\;
$(p_h,q_h) := getBoundaries(T, pos_h)$ \;
$res3 := buildIndexForAllGaps(q_l,p_h)$ \;
$res4 := insertDataBetweenBoundaries(pos_l, pos_h)$\;
$finalRes := res1 \cup res2 \cup res3 \cup res4$ \;
\Return $finalRes$\;
\end{algorithm}

This 
scenario is depicted in Fig. \ref{fig:different_partition_2} and the steps are illustrated in Algorithm \ref{algo:getResultsFromDifferentBound}. 
Let $l$ and $h$ belong to two different sorted partitions $S_l$ and $S_h$, which in our example are $S_2$ and $S_4$. Since both the query endpoints belong to two different sorted partitions, we can retrieve all entries residing in $S_l$ \textbf{(Lines 1--2)} and $S_h$ \textbf{(Lines 3--4)}. We also try to find all $gaps$  between $S_l$ and $S_h$, sort them and build learned index over them, and finally add all the necessary information to $T$ \textbf{(Lines 5--7)}. Then we try to find all entries in $PI$ that lie within partition boundaries \textbf{(Line 8)}. Finally, all the intermediate results are combined and returned. Here, $U_3$ and $U_4$ are the $gaps$ between $S_2$ and $S_4$, which are converted into $S_6$ and $S_7$.

\subsection{Case 4: Only $l$ belongs to an already sorted partition}

\begin{figure}[t]
\vspace{-12pt}
\begin{center}
\includegraphics[width=0.7\linewidth]{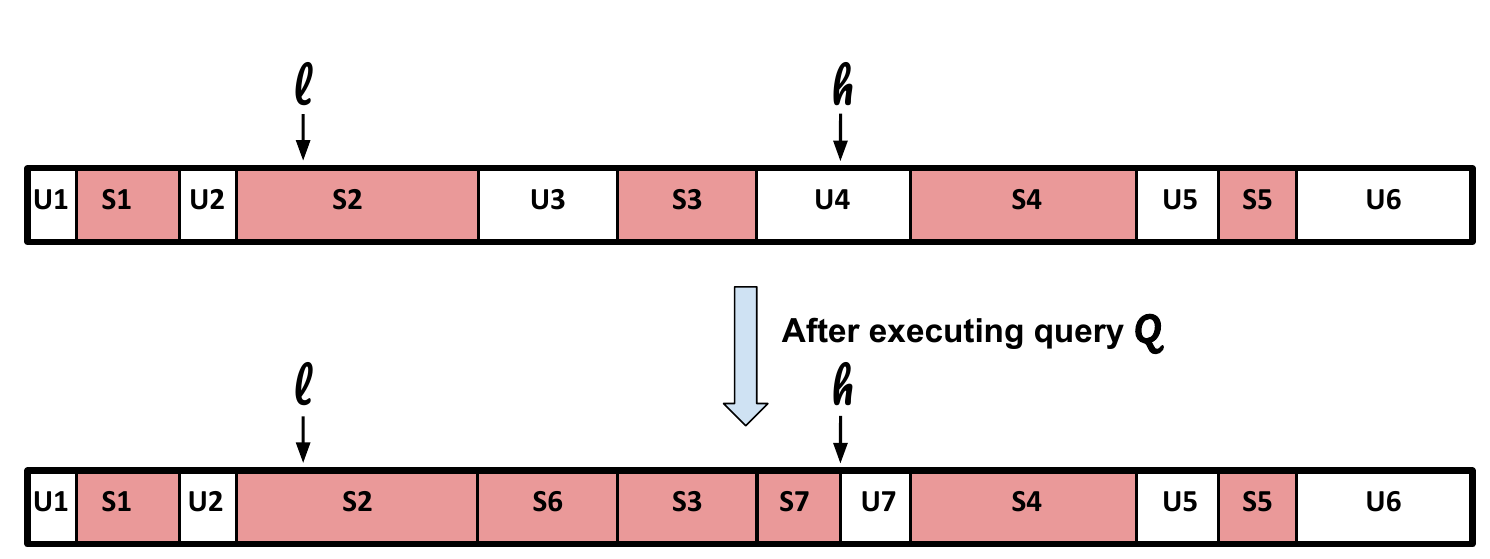}
\caption{Only $l$ belongs to an already sorted partition}
\label{fig:crack_high_2}
\end{center}
\end{figure}


\begin{algorithm}[t]
\caption{crackForHighValue$(l, h, pos_l)$}
\label{algo:crackForHighValue}
  \KwIn{$l,h$ are query boundaries. $pos_l$ is the index of partition in $T$ in which $l$ falls.
  }
  \KwOut{$finalResult$ satisfying the query boundary}
$M_l := getLearnedIndex(T, pos_l)$ //Get the learned index for $l$\;

$l_{pos} := find(M_l,l)$ //Get the position of $l$ from $M_l$\;
$tempPos := getHighValuePos(T, pos_l)$ \;
$res1 := A[l_{pos}:tempPos]$ \;
$resfromPI := findInPendingInserts(l,A[tempPos])$\;
$updateModel(M_l, resfromPI)$ \;
$(p_l,q_l) := getBoundaries(T, pos_l)$ //Get boundaries with $l$\;
$(p_h,q_h) := searchForGap(T,h)$ //Get the indices $U_h$\;
$res2 := buildIndexForAllGaps(q_l,p_h)$ \;
$minVal = min(A[p_h:q_h])$ //Get minimum value in $U_h$\;
$res3 := buildIndex(minVal,h)$ //Build index\;
$res4 := insertDataBetweenBoundaries(pos_l, q_h)$ \;
$finalRes := res1 \cup resfromPI \cup res2 \cup res3 \cup res4$ \;
\Return $finalRes$\;
\end{algorithm}

This scenario is depicted in Fig. \ref{fig:crack_high_2} and Algorithm \ref{algo:crackForHighValue} describes the steps. Using $T$, we can determine the sorted and unsorted partitions where $l$ and $h$ reside. Since $l$ belongs to a sorted partition, we use the corresponding learned index to 
identify the values falling within this partition satisfying the query \textbf{(Lines 1--4)}. Next, we identify results from $PI$ and update the learned model \textbf{(Lines 5--6)}. Similar to previous scenarios, we try to find all $gaps$, except the one that contains $h$, and then sort them, build learned indexes over them and update $T$ \textbf{(Lines 7--9)}.  We finally identify the $gap$ that contains $h$ and find the minimum value $minVal$ in that gap \textbf{(Line 10)}. Then, we build an index for $[minVal, h]$ \textbf{(Line 11)}. Then, we identify values from $PI$ falling in-between partition boundaries \textbf{(Line 12)}. Finally all intermediate results are combined to get the query result \textbf{(Lines 13--14)}.  In our example, $l$ lies in $S_2$, whereas $h$ belongs to $U_4$. $l_{pos}$ gets calculated by using the learned index of $S_2$. $U_4$ gets broken into the sorted partition $S_7$ and an unsorted partition $U_7$. The \textit{gap} $U_3$ gets converted to a sorted partition $S_6$ and gets added to $T$. 

\subsection{Case 5: Only $h$ belongs to an already sorted partition}

\begin{figure}[t]
\vspace{-12pt}
\begin{center}
\includegraphics[width=0.7\linewidth]{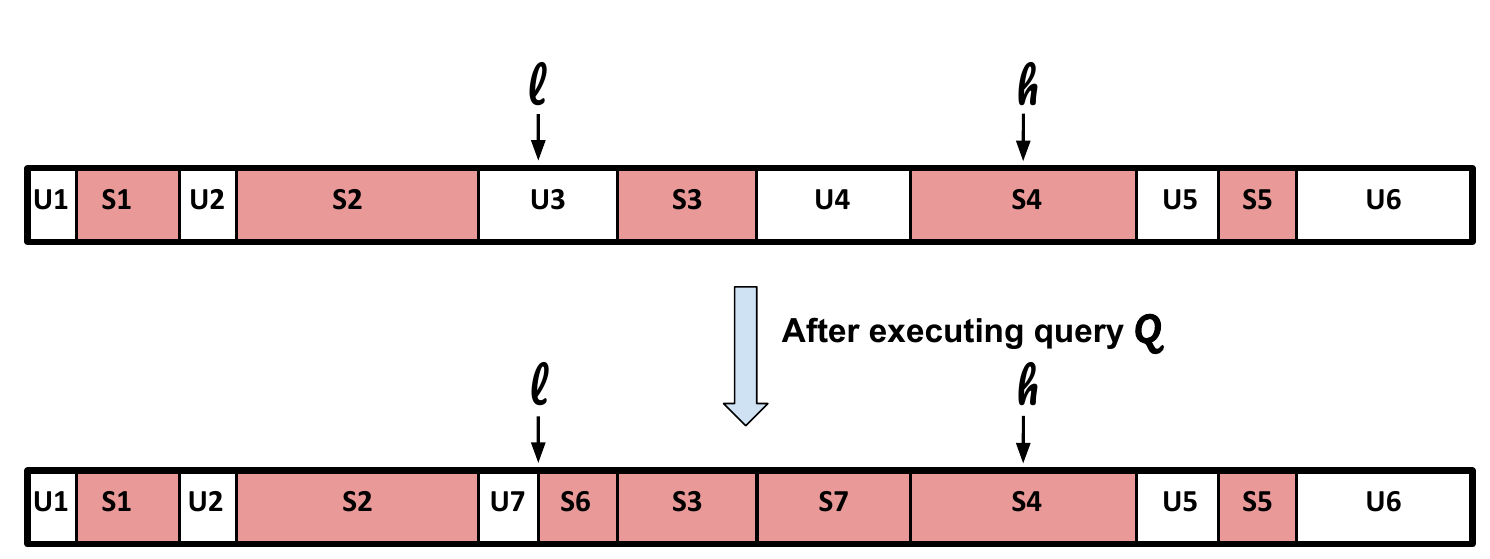}
\caption{Only $h$ belongs to an already sorted partition}
\label{fig:crack_low_2}
\end{center}
\end{figure}

This scenario is very similar to the previous scenario and is shown in Fig. \ref{fig:crack_low_2}. Here, instead of $h$, $l$ is cracked. The corresponding algorithm is skipped.
%

\section{Query Workload Prediction And Index Update}
\label{sec:query_prediction}

\begin{figure}[t]
\begin{flushleft}
  \begin{subfigure}[b]{0.45\linewidth}
	   \centering
	   \includegraphics[width=0.9\textwidth]{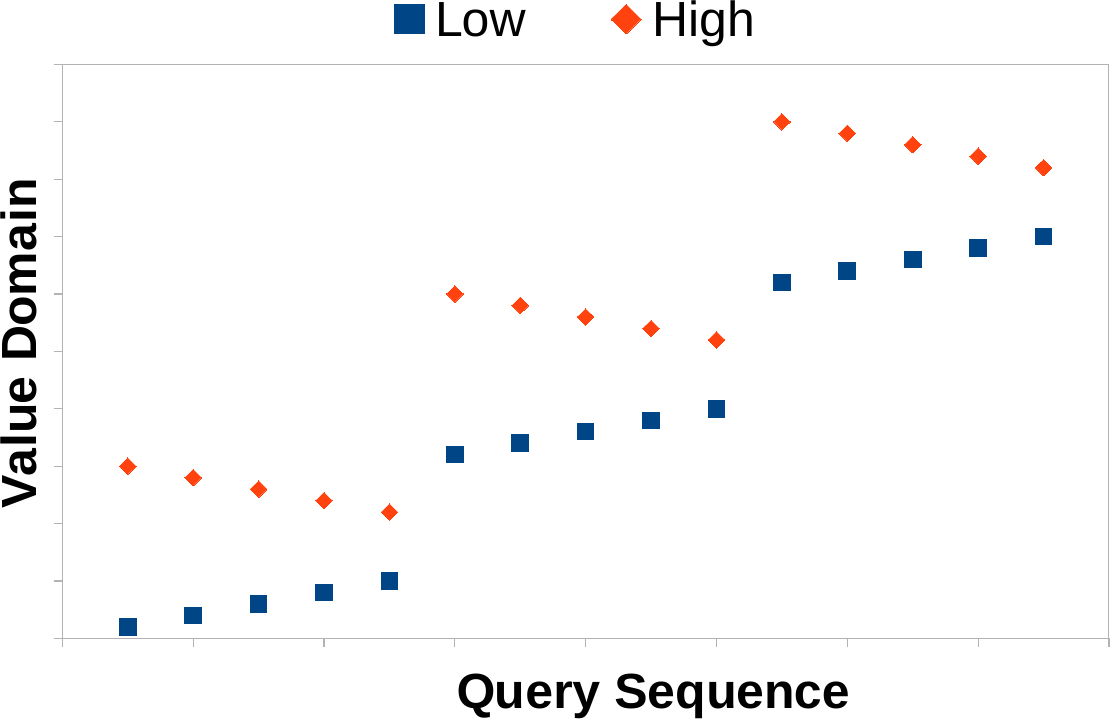}
	    \caption{Actual Queries}
	    \label{subfig:seq_zoomin_query_1}
	\end{subfigure}    
 \hspace{2pt}
  \begin{subfigure}[b]{0.45\linewidth}
	   \centering
	   \includegraphics[width=0.9\textwidth]{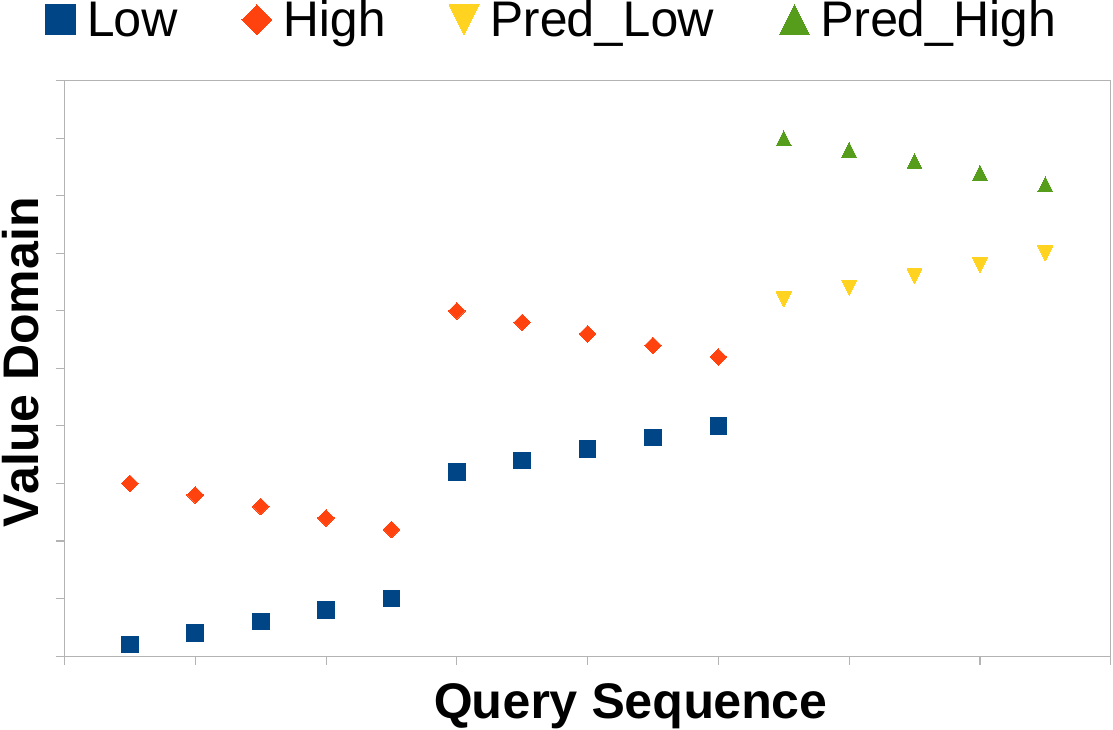}
	    \caption{Predicted Queries}
	    \label{subfig:seq_zoomin_pred}
	\end{subfigure}    
\caption{Sequential ZoomIn Queries}
\label{fig:seq_zoomin_query_and_pred}
\end{flushleft}
\end{figure}

In customer-centric use cases, there could be 
many scenarios, in which the queries often follow  specific patterns. In these cases, future queries are predictable from the previous queries, as shown in Fig. \ref{fig:seq_zoomin_query_and_pred}. 
The figure shows the query workload pattern in a \textit{Sequential ZoomIn} workload. In this workload, the queries zoom into a specific region of the data space and then shift to a different region. The zoom-in follows a linear trend both in terms of $low$ and $high$ boundaries, followed by a steady $jump$ to another region. Fig. \ref{subfig:seq_zoomin_query_1} shows the actual query ranges. If there are sufficient number of queries in the past, which maintains the $linearity$ \skd{(i.e. constant difference between consecutive $l$ or $h$ values, e.g. between 1$^{st}$ to 5$^{th}$ and again between 6$^{th}$ to 10$^{th}$ queries in Fig. \ref{fig:seq_zoomin_query_and_pred})} and the $jumps$ \skd{(i.e. sudden but constant change in $l$ or $h$ after  fixed set of queries, e.g. a constant jump between queries 5$^{th}$ and 6$^{th}$ and again between 10$^{th}$ and 11$^{th}$ query in Fig. \ref{fig:seq_zoomin_query_and_pred})}, then a future query workload can 
be predicted, as shown in Fig. \ref{subfig:seq_zoomin_pred}. Thus, \skd{when enough queries are executed, these can be utilized to `learn' the query patterns and forecast the next batch of queries by invoking a workload prediction algorithm in a separate thread. Since, modern CPUs support hundreds of cores, this would not affect the overall query performance.} 
The query ranges can be considered as a `time-series' data and so statistical methods for time series analysis and forecasting, such as ARIMA, can be employed for query workload forecasting. For each workload, the query patterns are different. Hence, it 
is challenging for a single prediction model to 
work for different workloads. 

\begin{algorithm}[t]
\caption{PredictWorkload$(\Delta, S, T)$}
\label{algo:PredictWorkload}
  \KwIn{$\Delta$ -- Past query batch \newline
   $S$ -- Set of forecasting methods \newline
   $T$ -- Existing Partition Table constructed so far 
  }
  \KwOut{Forecasted query batch $\Delta_f$}

$scores = \{\}$\;
\For {$M \in S$} {
    $MASE = M.apply(\Delta)$ //Get MASE for $M$\;
    $scores.append((M,MASE))$\;
}
$M_l$ = forecasting method having least $MASE$ in $scores$\;
$\Delta_f = M_l.forecast(\Delta)$ \;
$T.update(\Delta_f)$\;
\Return \;
\end{algorithm}

\skd{Our approach for query workload prediction is shown in Algorithm \ref{algo:PredictWorkload}.} In this, we maintain a fixed set of methods $S$ (such as ARIMA, AdaBoost, and Random Forest). To simplify matters, we assume that each workload batch contains $\Delta$ number of queries. 
After the execution of a workload batch, our \LAI{} structure is incrementally updated.
While the execution is in progress, the system launches the workload prediction task in a separate thread. In this task, every forecasting method $M \in S$ is applied over $\Delta$ \textbf{(Lines 1--4)}, after which the method ($M_l$) having the least MASE (Mean Absolute Scaled Error) is chosen \textbf{(Line 5)}. Then, $M_l$ is utilized to forecast the next batch of queries $\Delta_{f}$ \textbf{(Line 6)}. Once $\Delta_{f}$ is generated, it is used to further update \LAI{} based on the forecasted queries \textbf{(Line 7)}. In workloads, such as \textit{Sequential ZoomIn} of Fig. \ref{fig:seq_zoomin_query_and_pred}, the forecasts 
can be quite accurate due to the nature of the queries. Hence, when the next batch of queries is submitted, the query executor takes advantage of \LAI{} to answer the queries. It takes negligible amount of time to answer them, since the index is already built by $\Delta_{f}$ (\textbf{Case 2}). For workloads where the $\Delta_{f}$ is not very 
similar to the next batch of $\Delta$ queries, it can still be advantageous as it mostly falls in one of the cases except \textbf{Case 1(i)}. This process of query workload prediction and index updating 
continues as new batches of queries appear.
\begin{table}[]
\begin{tabular}{@{}|l|l|l|ll|@{}}
\toprule
\multicolumn{1}{|c|}{\multirow{2}{*}{\textbf{Dataset}}} & \multicolumn{1}{c|}{\multirow{2}{*}{\textbf{\begin{tabular}[c]{@{}c@{}}Size\\ (records)\end{tabular}}}} & \multicolumn{1}{c|}{\multirow{2}{*}{\textbf{Workloads}}}                                                                                                                                                           & \multicolumn{2}{c|}{\textbf{Operations}}                                  \\ \cmidrule(l){4-5} 
\multicolumn{1}{|c|}{}                                  & \multicolumn{1}{c|}{}                                                                                   & \multicolumn{1}{c|}{}                                                                                                                                                                                              & \multicolumn{1}{c|}{\textbf{Read}} & \multicolumn{1}{c|}{\textbf{Update}} \\ \midrule
Synthetic                                               & 100M                                                                                                    & \begin{tabular}[c]{@{}l@{}}Random, Sequential\\ Random, Sequential\\ Alternate, Sequential\\ Inverse, Sequential\\ Overlap, ZoomIn,\\ Sequential ZoomIn,\\ ZoomOut, Sequential\\ ZoomOut and Periodic\end{tabular} & \multicolumn{1}{l|}{100\%}         & 0\%                                  \\ \midrule
\multirow{3}{*}{YCSB}                                   & \multirow{3}{*}{100M}                                                                                   & \multirow{3}{*}{Uniform, Zipfian}                                                                                                                                                                                  & \multicolumn{1}{l|}{100\%}         & 0\%                                  \\ \cmidrule(l){4-5} 
                                                        &                                                                                                         &                                                                                                                                                                                                                    & \multicolumn{1}{l|}{90\%}          & 10\%                                 \\ \cmidrule(l){4-5} 
                                                        &                                                                                                         &                                                                                                                                                                                                                    & \multicolumn{1}{l|}{80\%}          & 20\%                                 \\ \bottomrule
\end{tabular}
\caption{Datasets and workloads}
\label{tab:workloads}
\end{table}

\section{Experimental Evaluation}
\label{sec:exp_evaluation}
In this section, we present our experimental evaluation of \LAI{}.

\subsection{Experimental Setting}

\LAI{} and \ULI{} were implemented in C++. The performance of \LAI{} was compared against other 
adaptive indexing techniques, such as database cracking (\textit{CRACK}) and stochastic cracking algorithms \cite{Stochastic_database_cracking}, such as \textit{DDC, DDR, DD1C, DD1R, MDD1R, AICC1R and AICS1R}. The C++ implementations of \textit{CRACK} and the other stochastic cracking algorithms were obtained from \cite{url:stochastic-cracking-impl}. 
All the implementations were compiled with \texttt{-O3} flag enabled. 
 The experiments were conducted on a system with Intel(R) Core(TM) i5-1035G1, having Ubuntu 24.04 LTS.

\begin{figure*}[!htbp]
  \begin{subfigure}[b]{0.3\textwidth}
	   \centering
	   \includegraphics[width=0.9\linewidth]{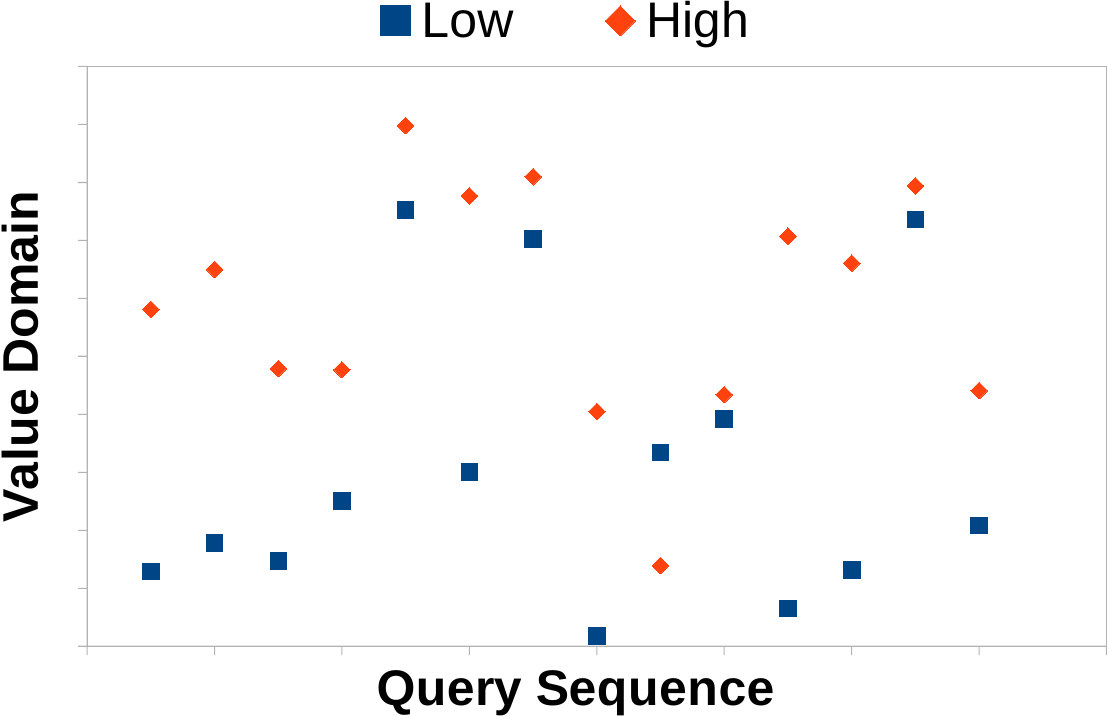}
	    \caption{Random}
	    \label{subfig:random_query}
	\end{subfigure}    
  \begin{subfigure}[b]{0.3\textwidth}
	   \centering
	   \includegraphics[width=0.9\linewidth]{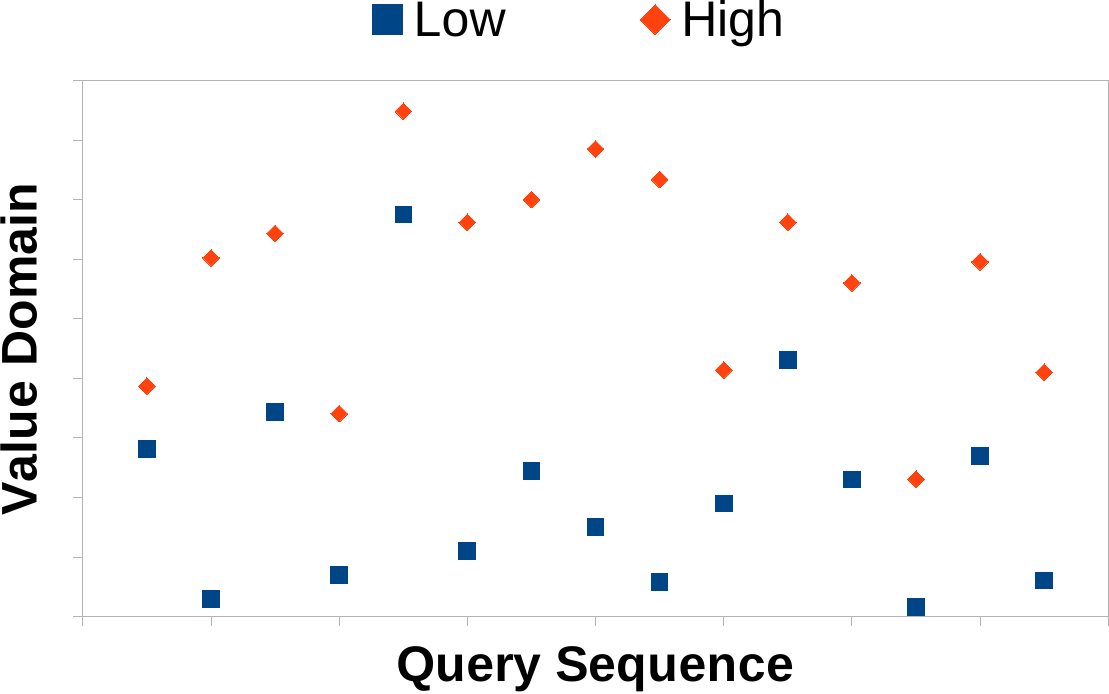}
	    \caption{Sequential Random}
	    \label{subfig:seq_rand_query}
	\end{subfigure}    
  \begin{subfigure}[b]{0.3\textwidth}
	   \centering
	   \includegraphics[width=0.9\linewidth]{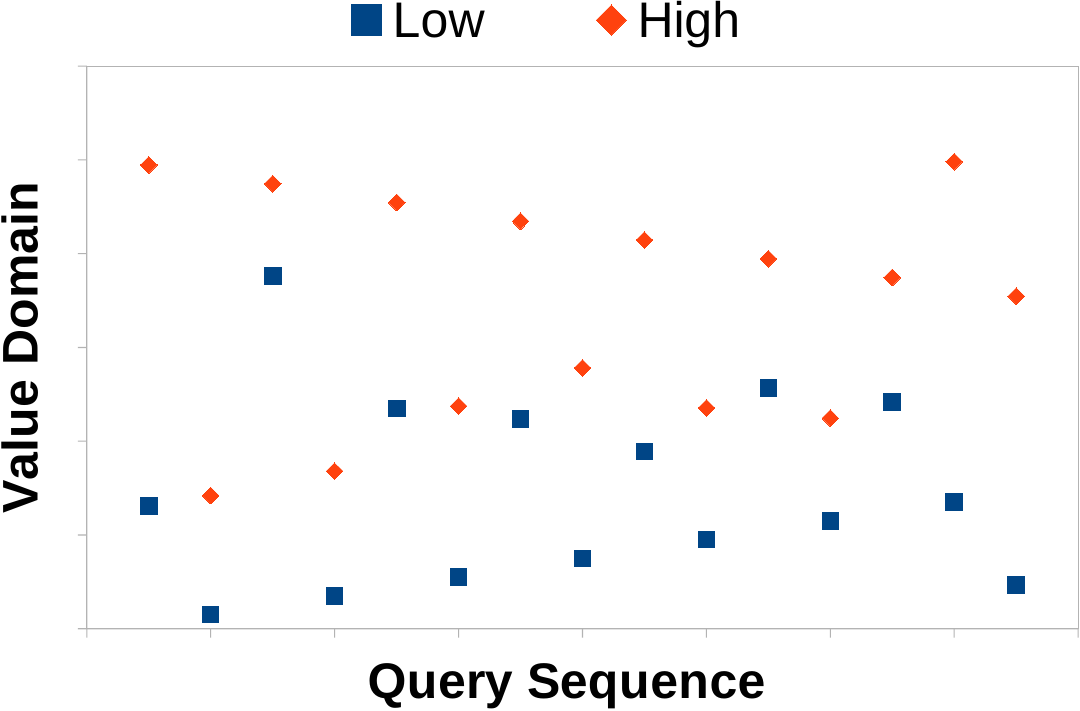}
	    \caption{Sequential Alternate}
	    \label{subfig:seq_alt_query}
	\end{subfigure}    
  \begin{subfigure}[b]{0.3\textwidth}
	   \centering
	   \includegraphics[width=0.9\linewidth]{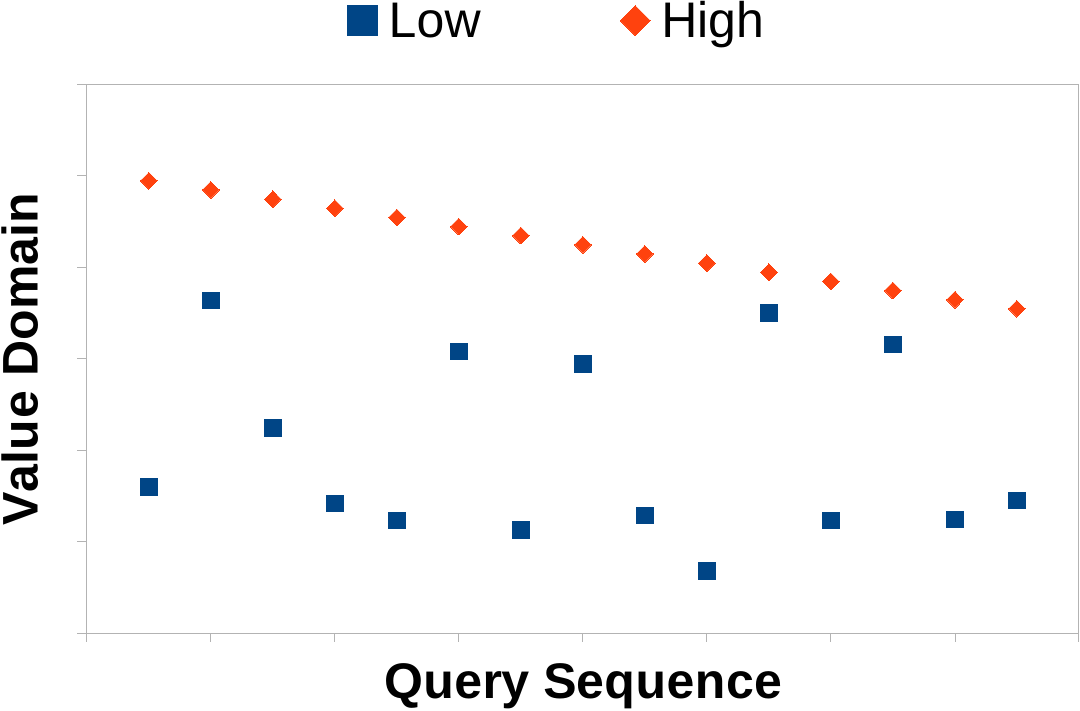}
	    \caption{Sequential Inverse}
	    \label{subfig:seq_inv_query}
	\end{subfigure}    
  \begin{subfigure}[b]{0.3\textwidth}
	   \centering
	   \includegraphics[width=0.9\linewidth]{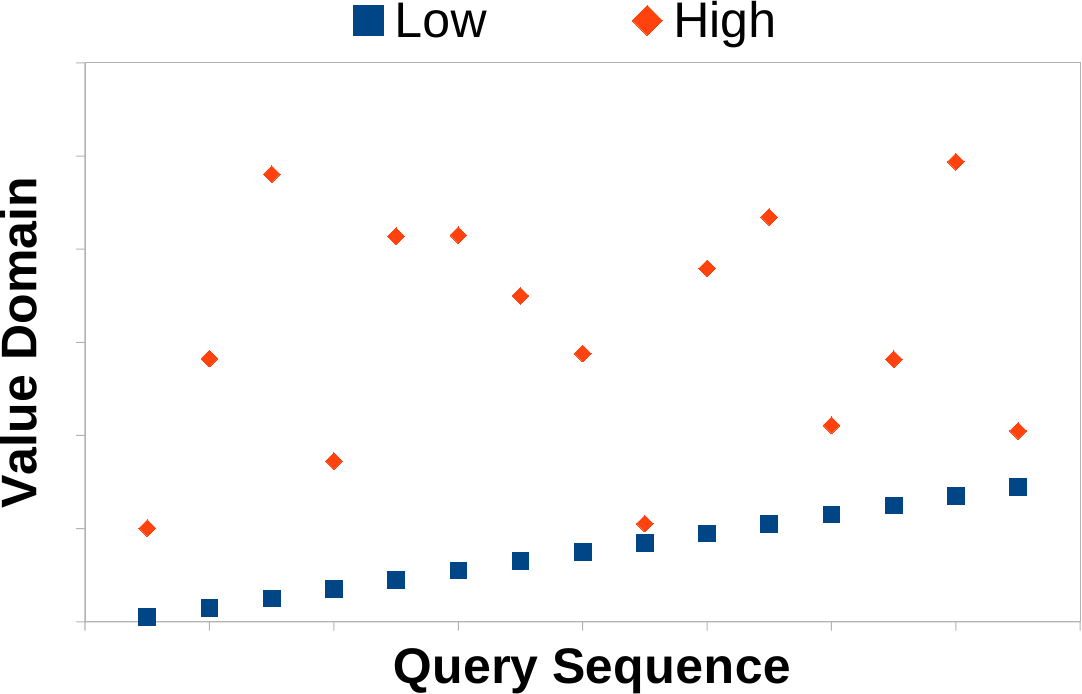}
	    \caption{Sequential Overlap}
	    \label{subfig:seq_over_query}
	\end{subfigure}    
  \begin{subfigure}[b]{0.3\textwidth}
	   \centering
	   \includegraphics[width=0.9\linewidth]{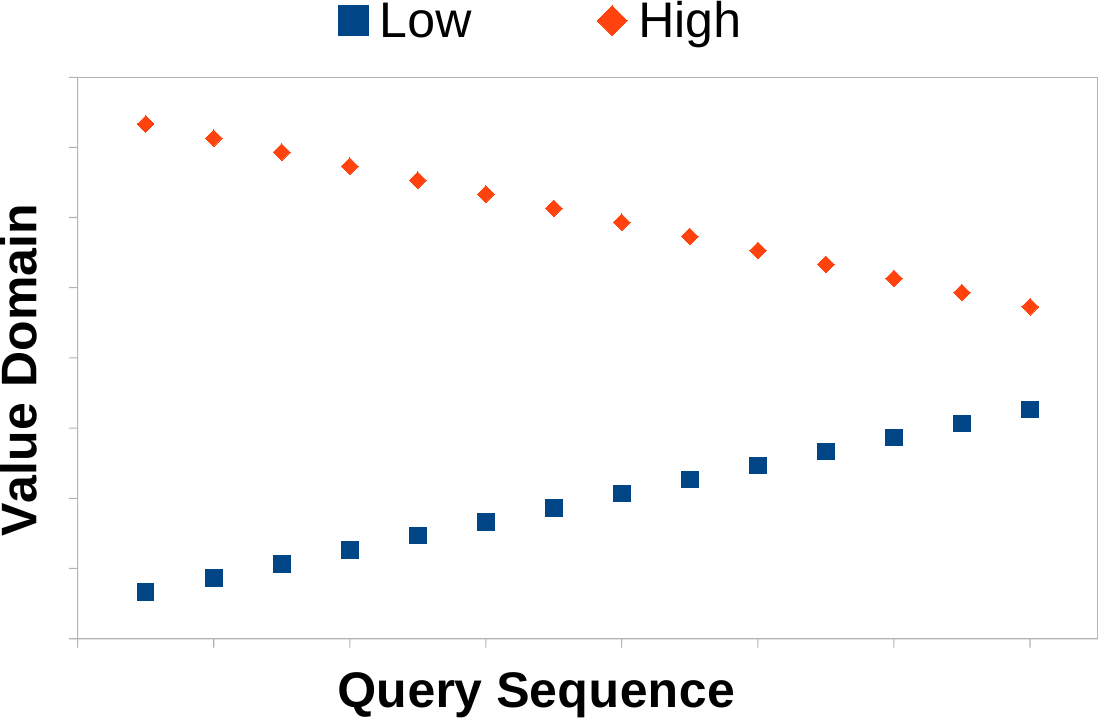}
	    \caption{ZoomIn}
	    \label{subfig:zoomin_query}
	\end{subfigure}    
  \begin{subfigure}[b]{0.3\textwidth}
	   \centering
	   \includegraphics[width=0.9\linewidth]{figures/seq_zoomin_query.pdf}
	    \caption{Sequential ZoomIn}
	    \label{subfig:seq_zoomin_query}
	\end{subfigure}    
  \begin{subfigure}[b]{0.3\textwidth}
	   \centering
	   \includegraphics[width=0.9\linewidth]{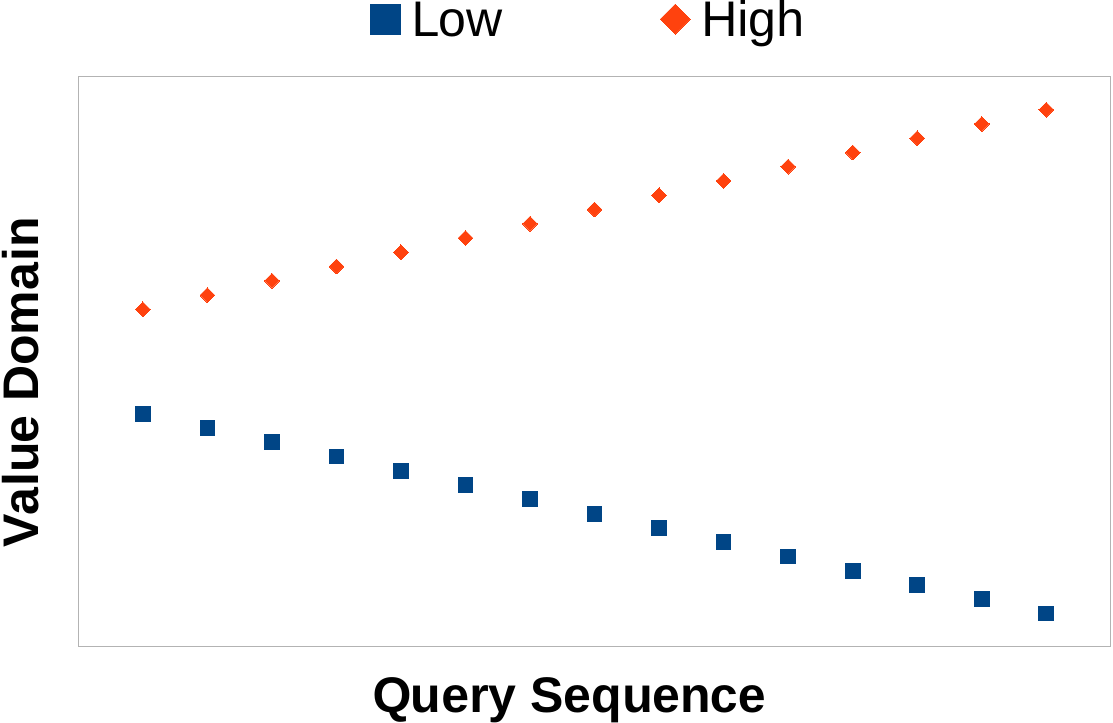}
	    \caption{ZoomOut}
	    \label{subfig:zoomout_query}
	\end{subfigure}    
  \begin{subfigure}[b]{0.3\textwidth}
	   \centering
	   \includegraphics[width=0.9\linewidth]{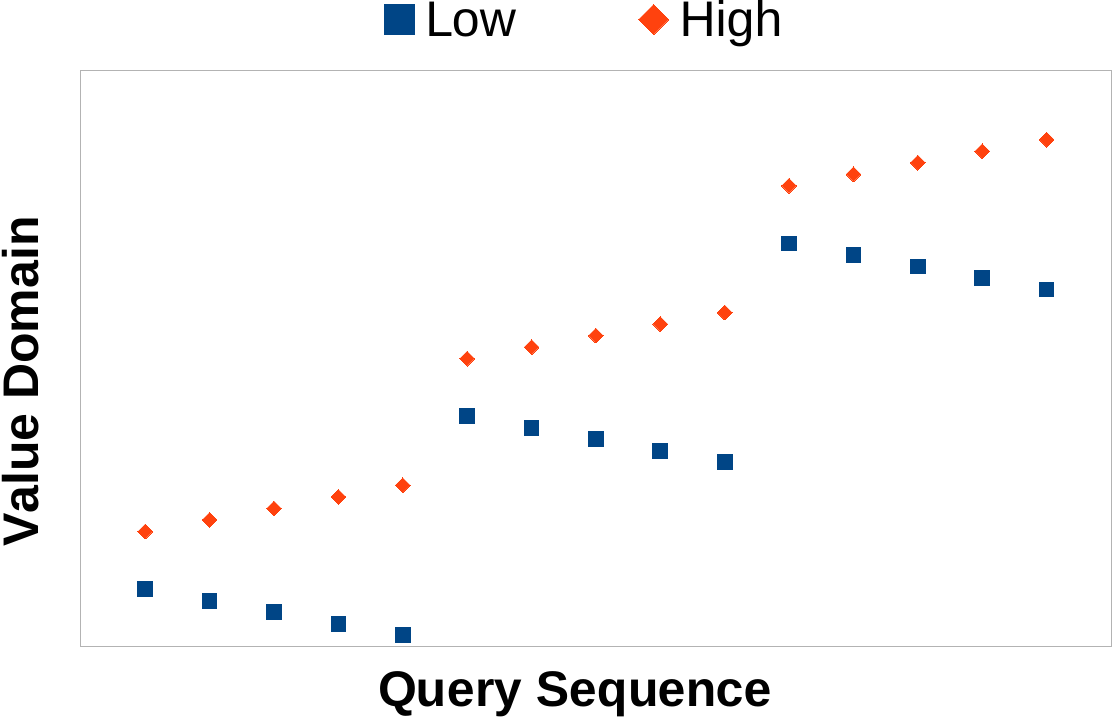}
	    \caption{Sequential ZoomOut}
	    \label{subfig:seq_zoomout_query}
	\end{subfigure}    
  \begin{subfigure}[b]{0.3\textwidth}
	   \centering
	   \includegraphics[width=0.9\linewidth]{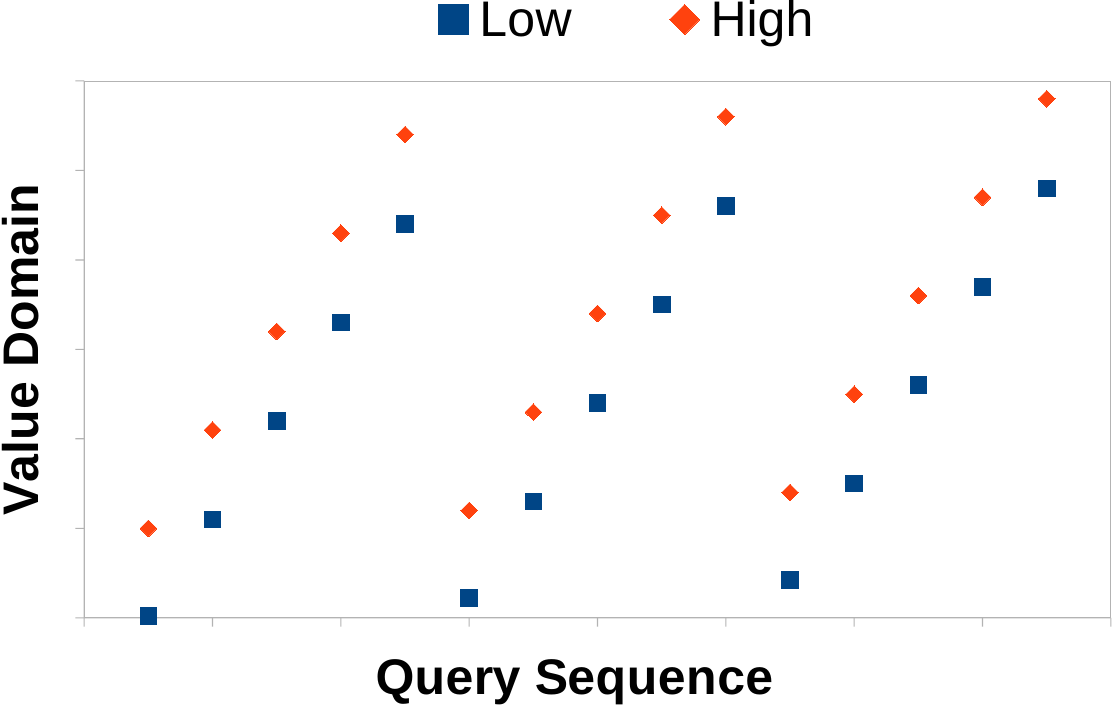}
	    \caption{Periodic}
	    \label{subfig:periodic_query}
	\end{subfigure}    
\caption{Synthetic query workloads}
\label{fig:synthetic_query_workloads}
\end{figure*}

\begin{figure*}[!htbp]
  \begin{subfigure}[b]{\textwidth}
	   \centering
	\includegraphics[width=\linewidth, trim=0 0 0 285pt,
    clip]{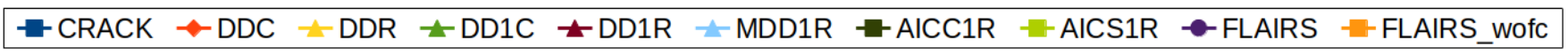}
	    \label{subfig:eval_labels}
    \vspace{-180pt}
	\end{subfigure}  
  \begin{subfigure}[b]{0.3\textwidth}
	   \centering
	   \includegraphics[width=\linewidth]{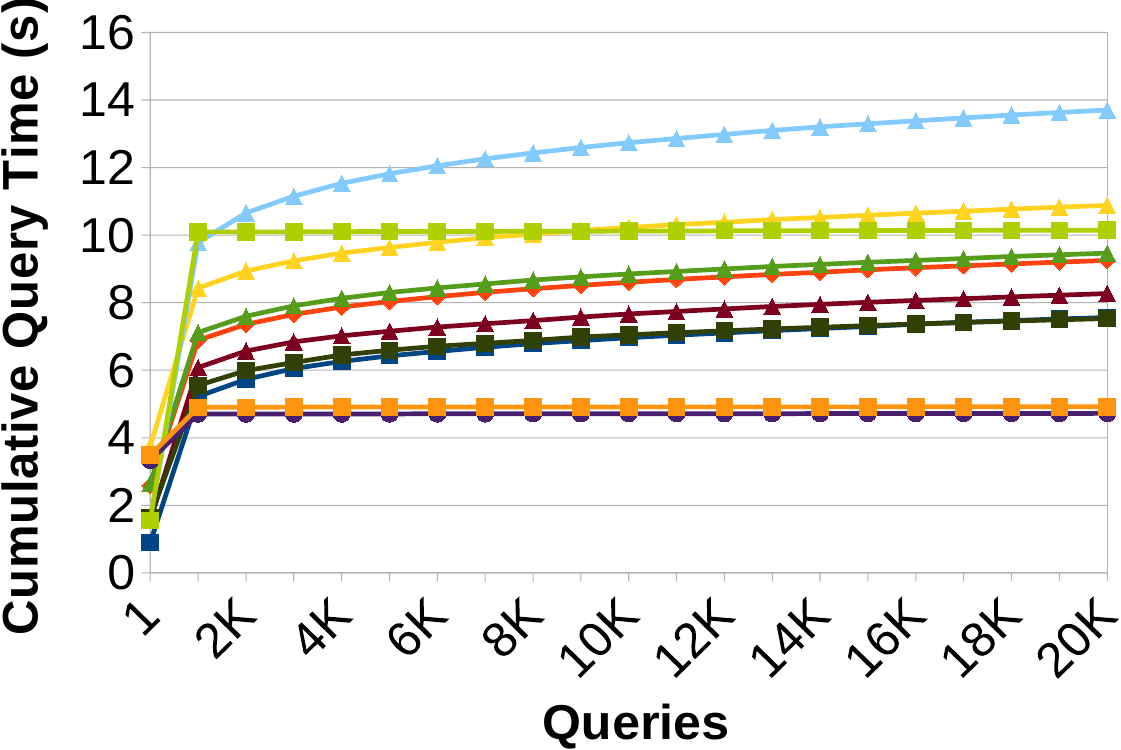}
	    \caption{Random}
	    \label{subfig:eval_random}
	\end{subfigure}    
  \begin{subfigure}[b]{0.3\textwidth}
	   \centering
	   \includegraphics[width=\linewidth]{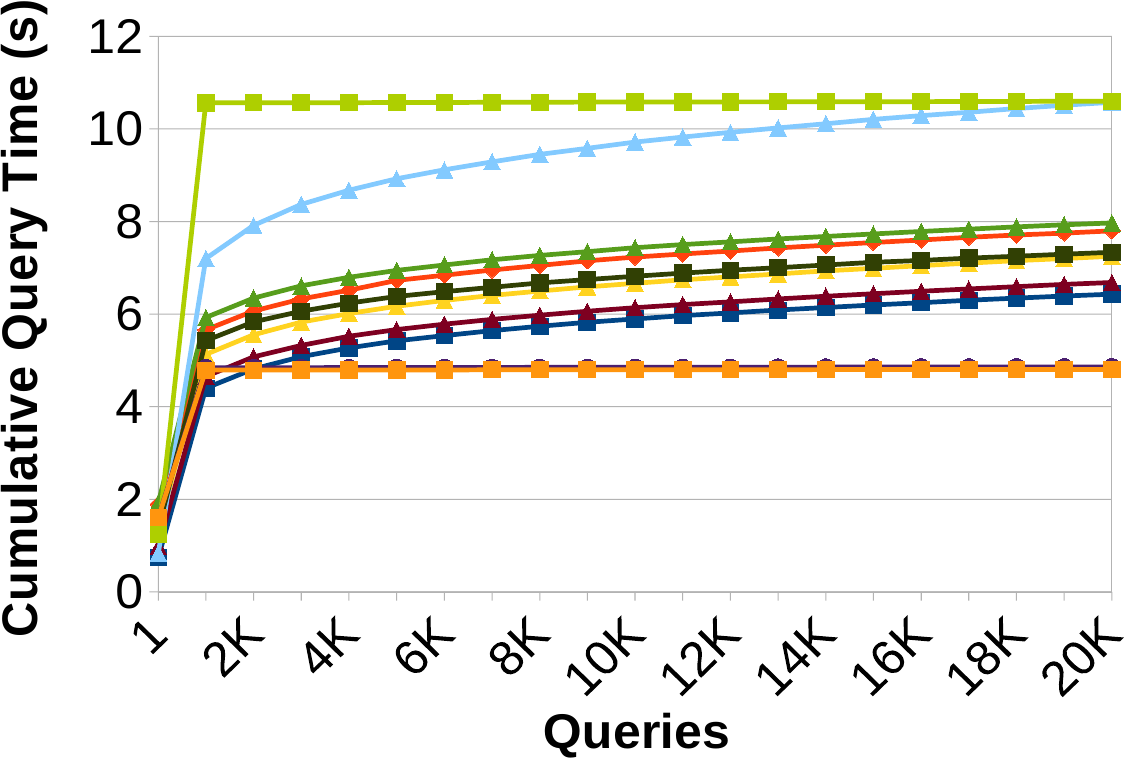}
	    \caption{Sequential Random}
	    \label{subfig:eval_seq_rand}
	\end{subfigure}    
  \begin{subfigure}[b]{0.3\textwidth}
	   \centering
	   \includegraphics[width=\linewidth]{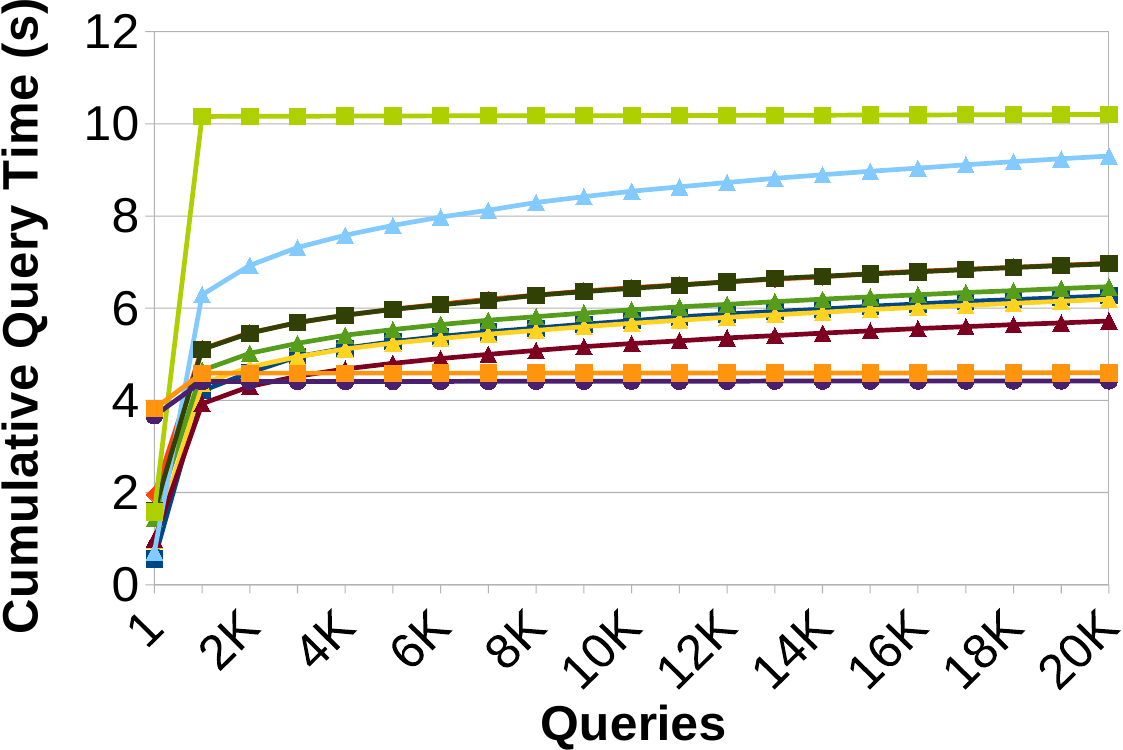}
	    \caption{Sequential Alternate}
	    \label{subfig:eval_seq_alt}
	\end{subfigure}    
  \begin{subfigure}[b]{0.3\textwidth}
	   \centering
	   \includegraphics[width=\linewidth]{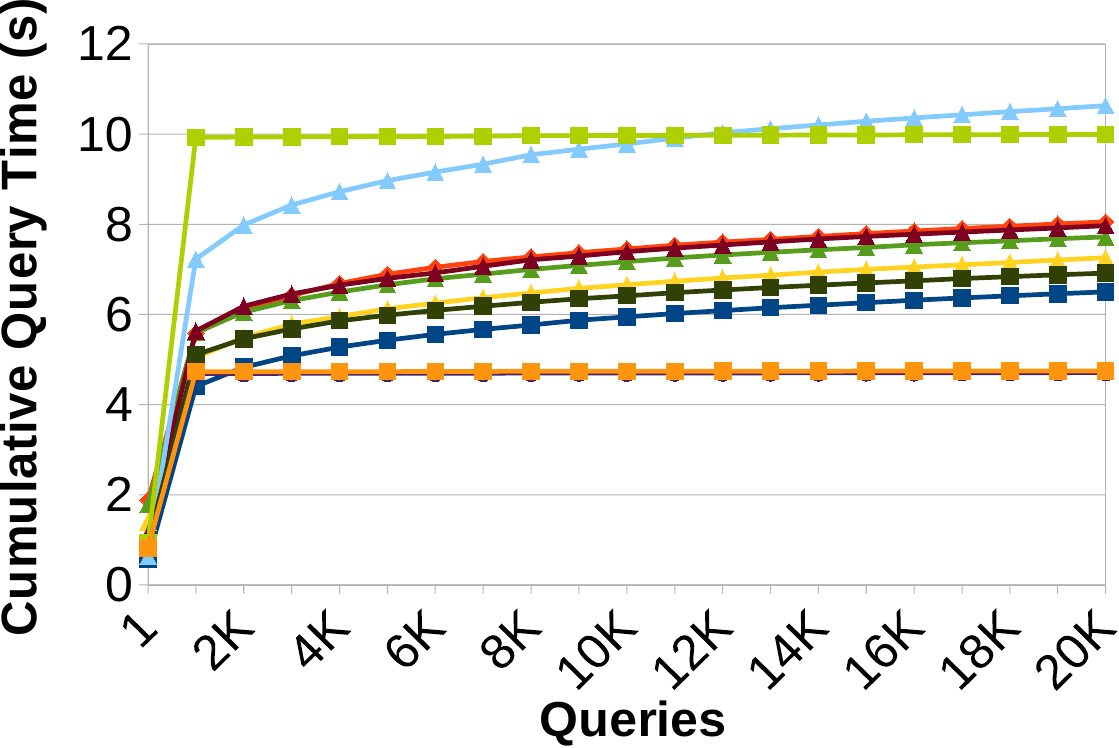}
	    \caption{Sequential Inverse}
	    \label{subfig:eval_seq_inv}
	\end{subfigure}    
  \begin{subfigure}[b]{0.3\textwidth}
	   \centering
	   \includegraphics[width=\linewidth]{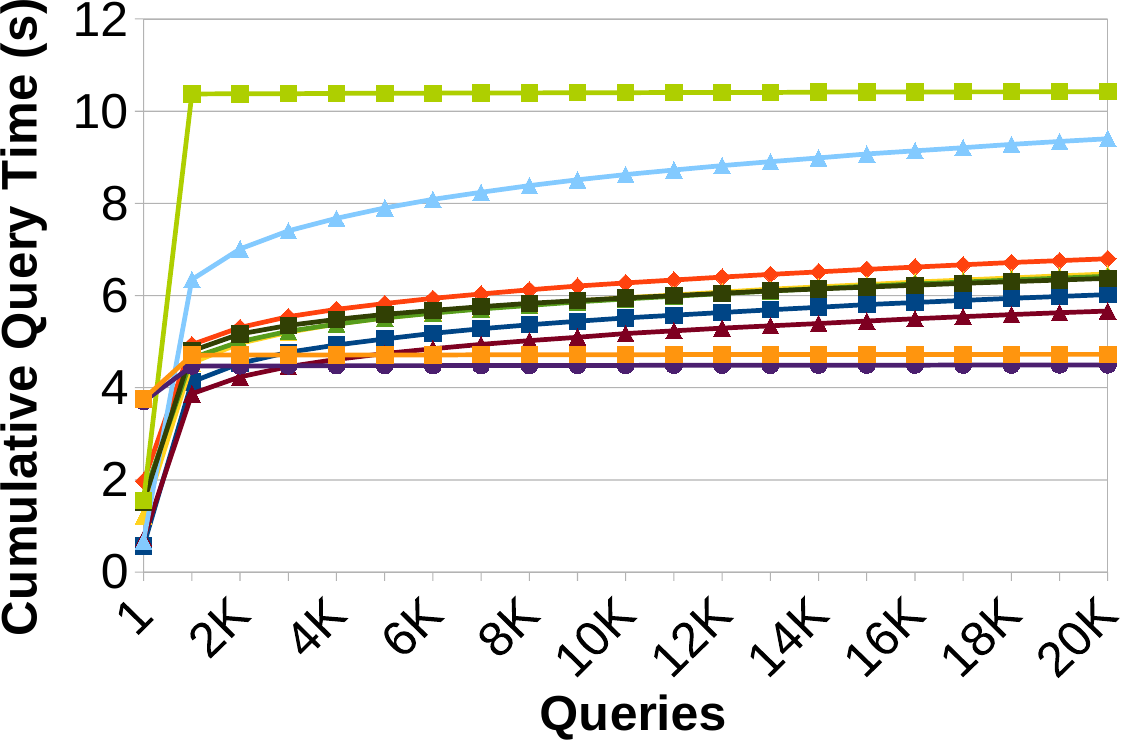}
	    \caption{Sequential Overlap}
	    \label{subfig:eval_seq_over}
	\end{subfigure}    
  \begin{subfigure}[b]{0.3\textwidth}
	   \centering
	   \includegraphics[width=\linewidth]{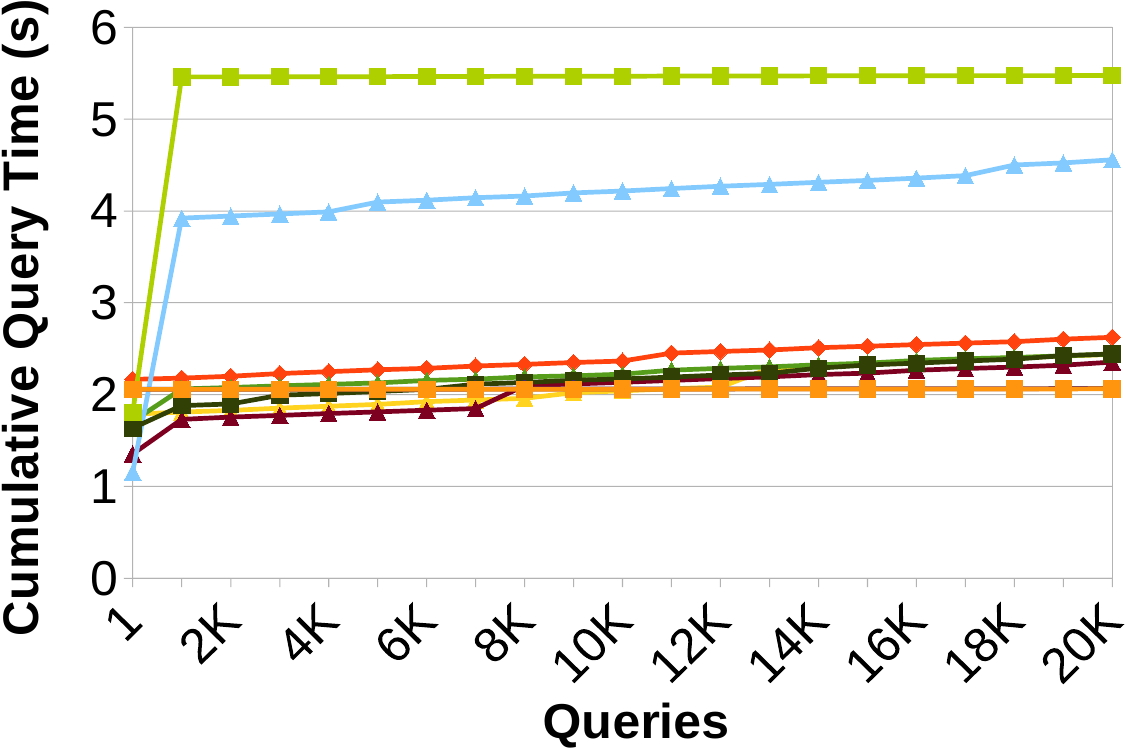}
	    \caption{ZoomIn}
	    \label{subfig:eval_zoomin}
	\end{subfigure}    
 \hfill
  \begin{subfigure}[b]{0.3\textwidth}
	   \centering
	   \includegraphics[width=\linewidth]{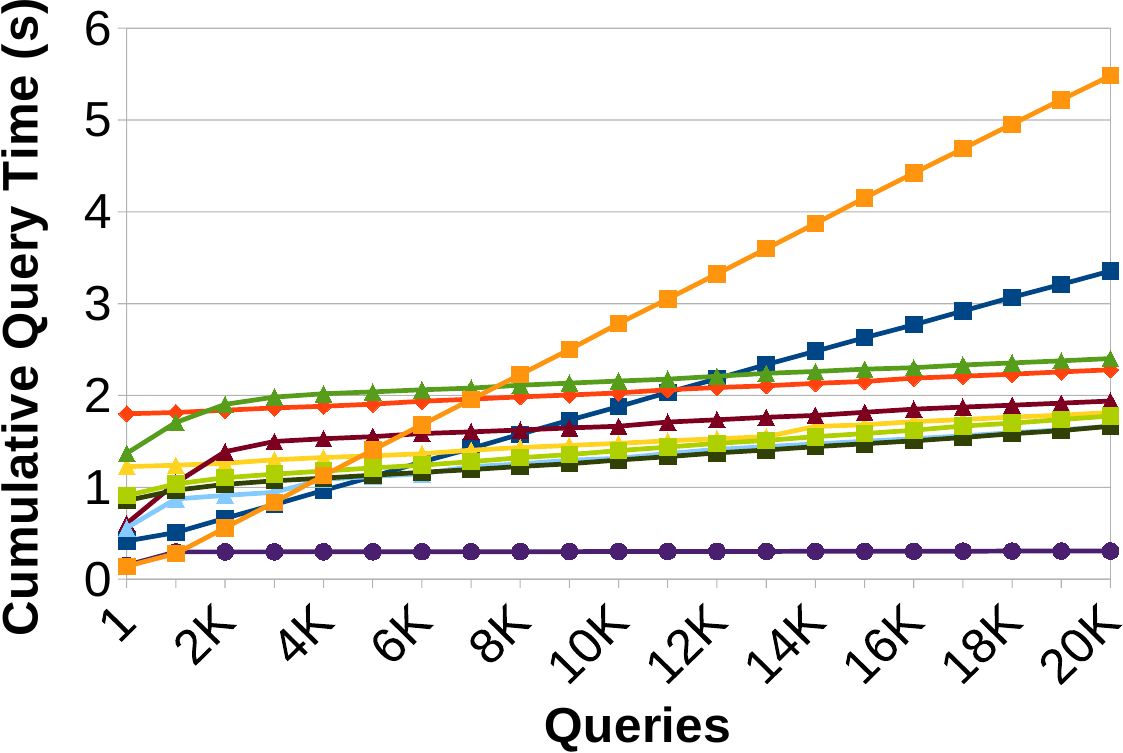}
	    \caption{Sequential ZoomIn}
	    \label{subfig:eval_seq_zoomin}
	\end{subfigure}    
  \begin{subfigure}[b]{0.3\textwidth}
	   \centering
	   \includegraphics[width=\linewidth]{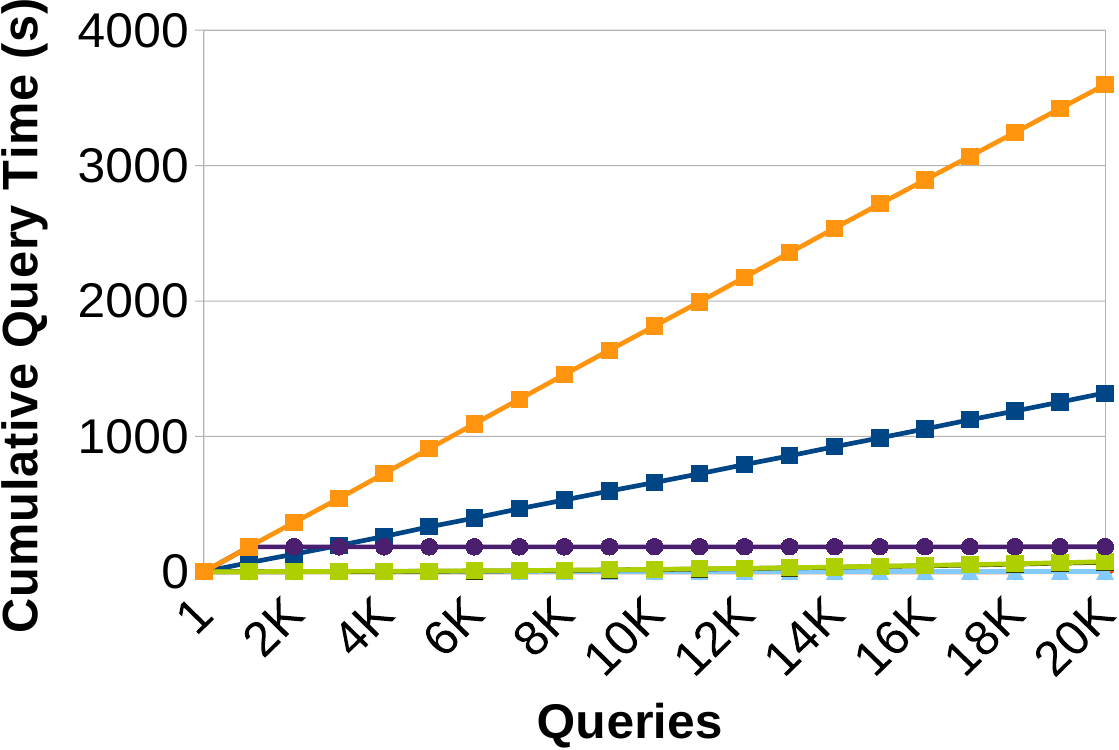}
	    \caption{ZoomOut}
	    \label{subfig:eval_zoomout}
	\end{subfigure}    
  \begin{subfigure}[b]{0.3\textwidth}
	   \centering
	   \includegraphics[width=\linewidth]{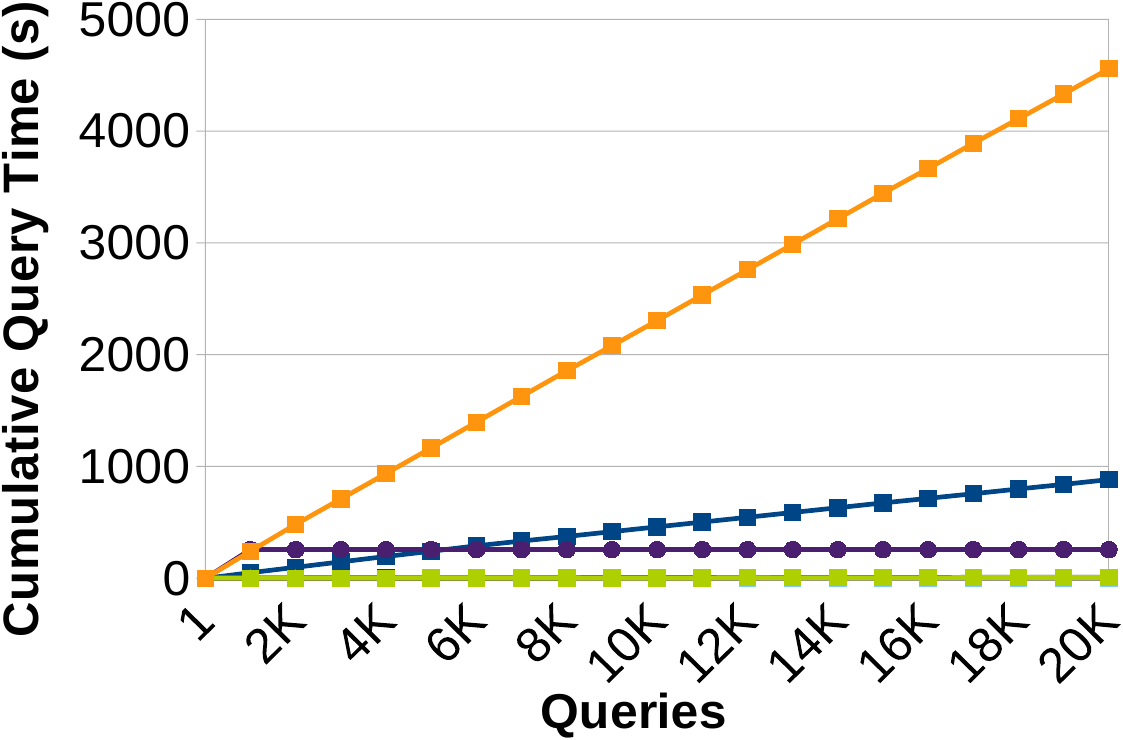}
	    \caption{Sequential ZoomOut}
	    \label{subfig:eval_seq_zoomout}
	\end{subfigure}    
  \begin{subfigure}[b]{0.3\textwidth}
	   \centering
	   \includegraphics[width=\linewidth]{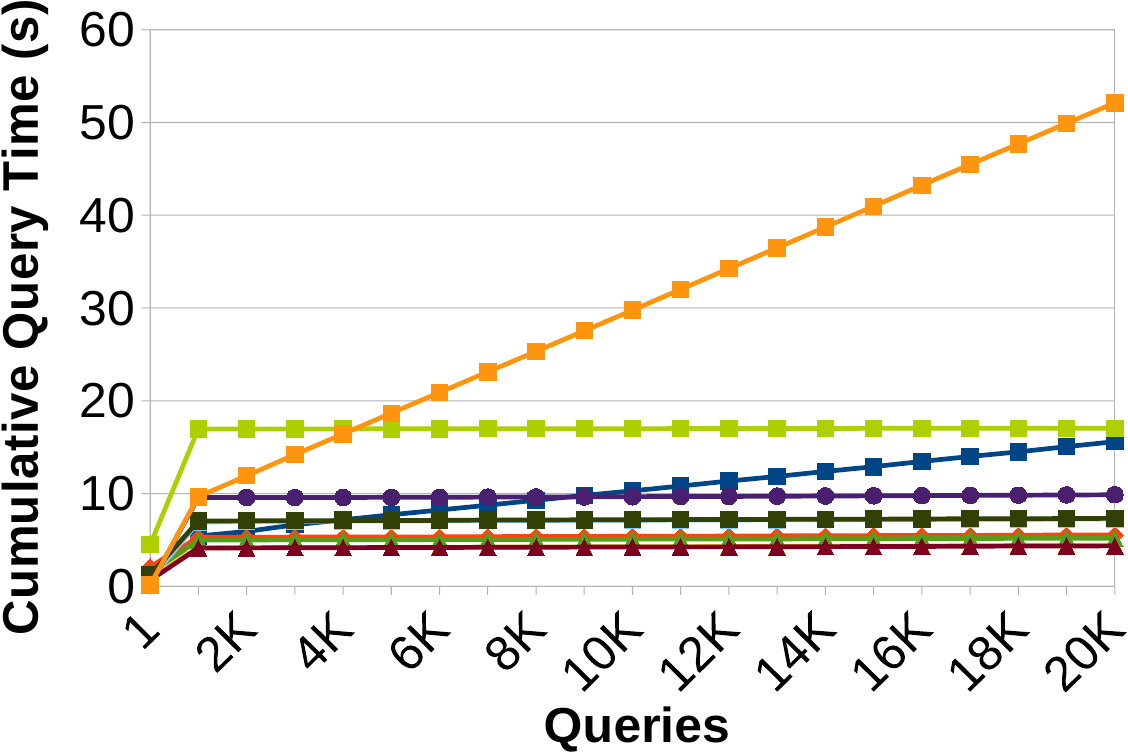}
	    \caption{Periodic}
	    \label{subfig:eval_periodic}
	\end{subfigure}    
\caption{Evaluation on synthetic workloads}
\label{fig:eval_synthetic_query_workloads}
\vspace{-12pt}
\end{figure*}

\subsection{Datasets and workloads} 
\label{sec:dataset_and_workloads}
To evaluate the performance of \LAI, both a synthetic dataset and a YCSB~\cite{Cooper2010YCSB} benchmark dataset are used. 
The details of the datasets and workloads are in Table~\ref{tab:workloads}.
Formally, queries were executed on a randomly shuffled dataset \textit{A} having all keys between [0, 100M). The queries were of the form: \textit{Find all $x \in$ A such that $l \leq x \leq h$}. The queries were generated from ten different  workloads for the synthetic dataset
and from two   workloads for the YCSB dataset. 

An illustration of query  workloads in the synthetic dataset is shown in Fig. \ref{fig:synthetic_query_workloads}.  It can be observed that for \textit{Random}, both $l$ and $h$ were generated randomly. In \textit{Sequential Random}, for the odd numbered queries $l$ was random, whereas for the even ones $l$ was sequentially increased. However, $h$ was still random. In \textit{Sequential Alternate}, for the odd numbered queries $l$ was randomly generated but $h$ was gradually decreased and vice-versa (i.e. $l$ increased but $h$ was random) for the even numbered queries. For \textit{Sequential Inverse}, $h$ sequentially decreased, but $l$ was random. \textit{Sequential Overlap} is the reverse of \textit{Sequential Inverse}, where $l$ sequentially increased but $h$ was random.
The query patterns for \textit{ZoomIn}, \textit{Sequential ZoomIn}, \textit{ZoomOut} and \textit{Sequential ZoomOut} are easily identifiable from their respective figures.
In \textit{Periodic}, the difference between \textit{l} and \textit{h} remains constant, but the query boundary changes.

\subsection{Experimental Results}
We present our evaluation of \LAI{} against other adaptive indexing approaches, with both synthetic and YCSB workloads.

\subsubsection{\textbf{Evaluating \LAI{} with other approaches on synthetic workloads.}}

The results of the experiments are shown in Fig. \ref{fig:eval_synthetic_query_workloads}. From the figure we see that the performance of both \LAI{} and \LAI$_{wofc}$ (\LAI{} \textbf{w}ith\textbf{o}ut query \textbf{f}ore\textbf{c}asting), in-terms of cumulative query execution time, is almost the same for \textit{Random, Sequential Random, Sequential Alternate, Sequential Inverse, Sequential Overlap,} and \textit{ZoomIn} queries (Figs. \ref{subfig:eval_random} -- \ref{subfig:eval_zoomin}). But there is a significant improvement in the cumulative query execution time of \LAI{} compared to \LAI$_{wofc}$ for the remaining four workloads (Figs. \ref{subfig:eval_seq_zoomin} -- \ref{subfig:eval_periodic}).  Although the execution times for the first query in both \LAI{} and \LAI$_{wofc}$ are high, but as more queries are submitted, their performance starts to improve. For \LAI{} the cumulative query execution time almost remains constant after 1K queries across all workloads, but for \LAI$_{wofc}$, it keeps linearly increasing for \textit{Sequential ZoomIn, ZoomOut, Sequential ZoomOut} and \textit{Periodic} workloads (Figs. \ref{subfig:eval_seq_zoomin} -- \ref{subfig:eval_periodic}) after 1K queries. This suggests that the cumulative query execution time is dominated by the first 1K queries in case of \LAI, after which the execution time remains constant due to query forecasting, especially for the last four workloads (Figs. \ref{subfig:eval_seq_zoomin} -- \ref{subfig:eval_periodic}). \sdn{The almost constant cumulative execution time after a certain number of queries is an indication that the index has been built on the entire dataset, after those queries got executed.}

\begin{figure*}[htbp]
  \begin{subfigure}[b]{0.47\textwidth}
	   \centering
	   \includegraphics[width=\linewidth]{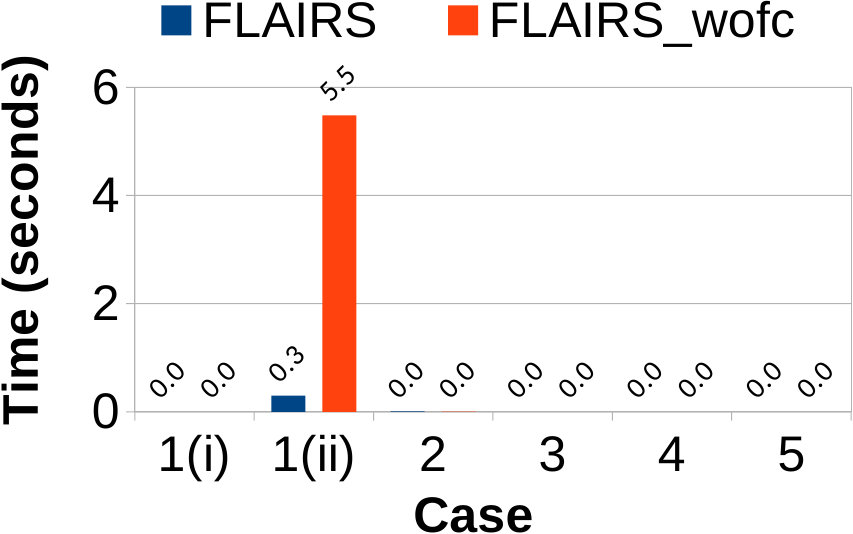}
	    \caption{Sequential ZoomIn - Time}
	    \label{subfig:eval_seq_zoomin_case_time}
	\end{subfigure}    
  \begin{subfigure}[b]{0.47\textwidth}
	   \centering
	   \includegraphics[width=\linewidth]{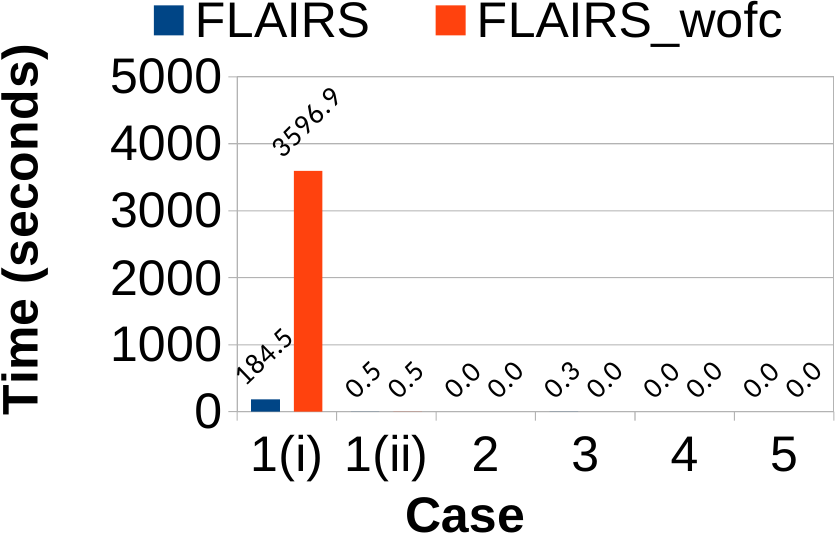}
	    \caption{ZoomOut - Time}
	    \label{subfig:eval_zoomout_case_time}
	\end{subfigure}    
  \begin{subfigure}[b]{0.47\textwidth}
	   \centering
	   \includegraphics[width=\linewidth]{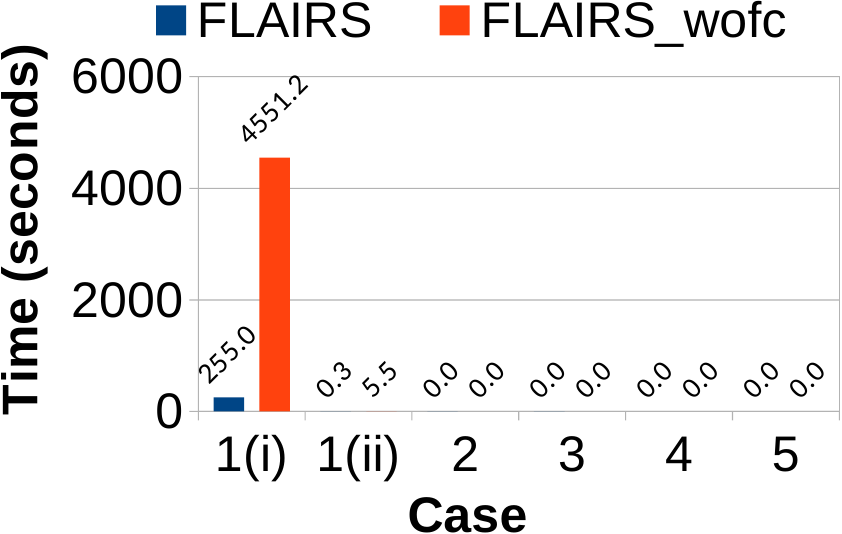}
	    \caption{Sequential ZoomOut - Time}
	    \label{subfig:eval_seq_zoomout_case_time}
	\end{subfigure}    
 \hspace{15pt}
  \begin{subfigure}[b]{0.47\textwidth}
	   \centering
	   \includegraphics[width=\linewidth]{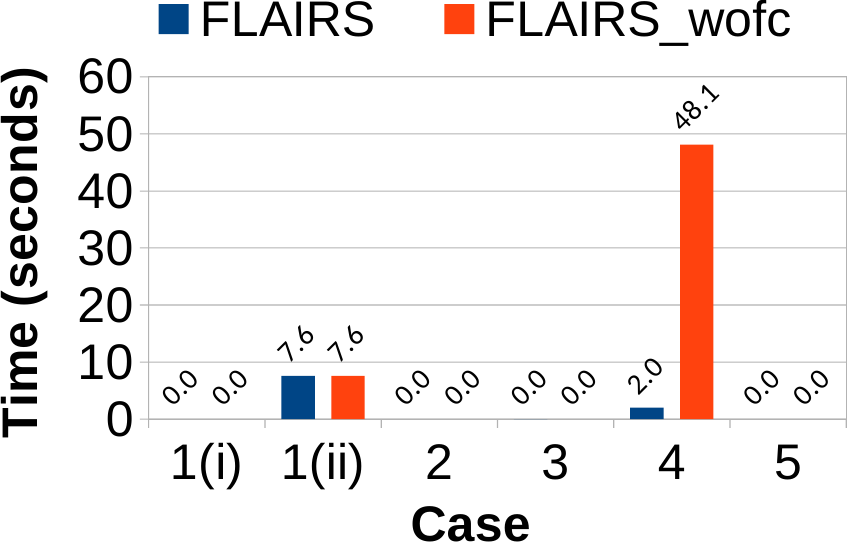}
	    \caption{Periodic - Time}
	    \label{subfig:eval_periodic_case_time}
	\end{subfigure}    
\caption{Casewise evaluation of  
\textit{Sequential ZoomIn, ZoomOut, Sequential ZoomOut} and \textit{Periodic} in  synthetic workloads }
\label{fig:eval_synthetic_query_workloads_casewise}
\end{figure*}

\begin{figure*}[!t]
  \begin{subfigure}[b]{\textwidth}
	   \centering
	   \includegraphics[width=\linewidth, trim=0 0 0 285pt,
    clip]{figures/eval_labels.pdf}
	    \vspace{-180pt}
	\end{subfigure}    
 \vspace{0.5pt}
  \begin{subfigure}[b]{0.3\textwidth}
	   \centering
	   \includegraphics[width=\linewidth]{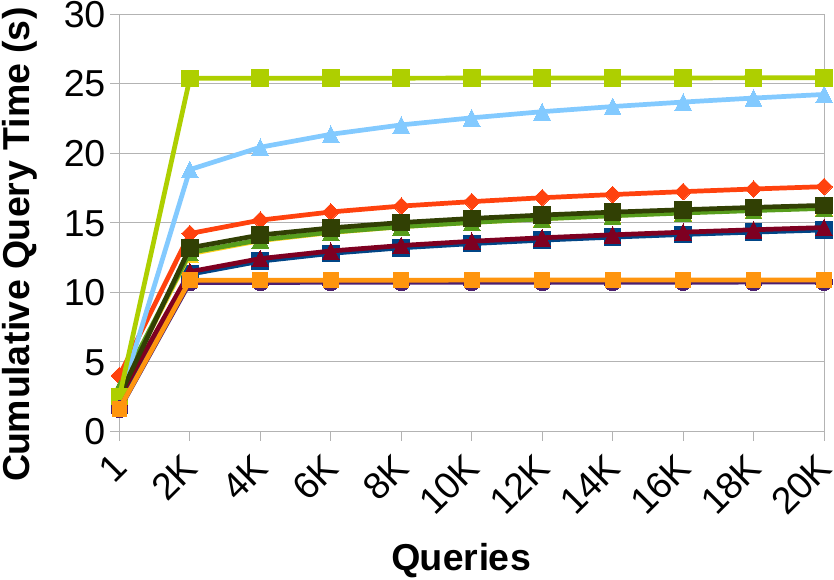}
	    \caption{Uniform (Read 100\%)}
	    \label{subfig:exp_uniform_read_only}
	\end{subfigure}    
  \begin{subfigure}[b]{0.3\textwidth}
	   \centering
	   \includegraphics[width=\linewidth]{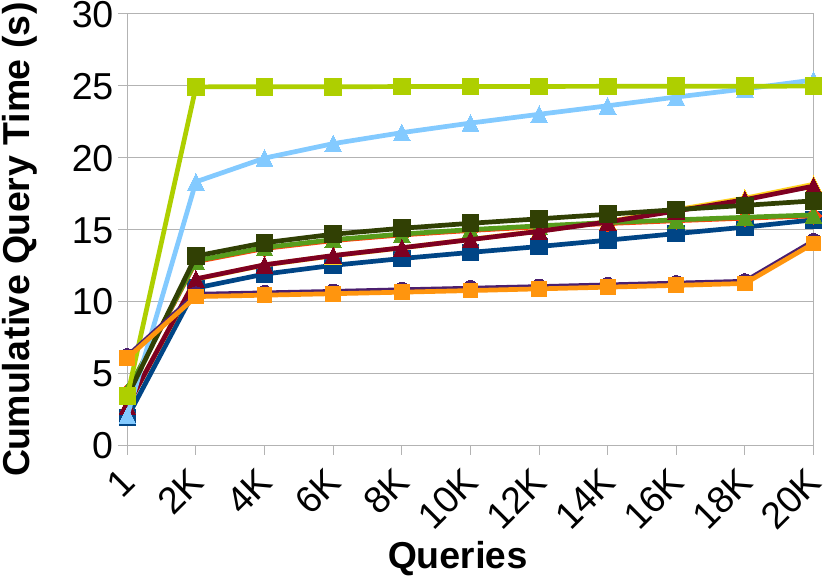}
	    \caption{Uniform (Read 90\%, Insert 10\%)}
	    \label{subfig:exp_uniform_read_90_ins_10}
	\end{subfigure}    
  \begin{subfigure}[b]{0.3\textwidth}
	   \centering
	   \includegraphics[width=\linewidth]{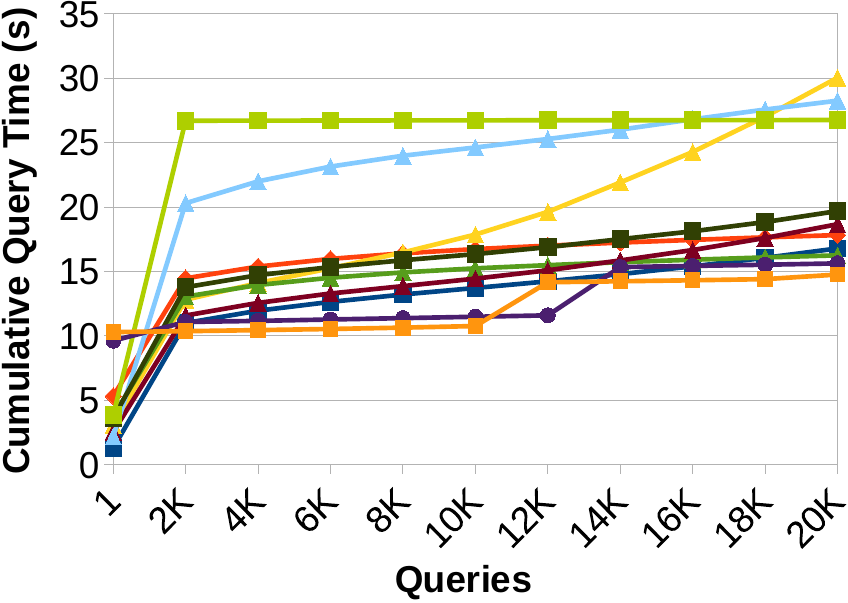}
	    \caption{Uniform (Read 80\%, Insert 20\%)}
	    \label{subfig:exp_uniform_read_80_insert_20}
	\end{subfigure}    
  \begin{subfigure}[b]{0.3\textwidth}
	   \centering
	   \includegraphics[width=\linewidth]{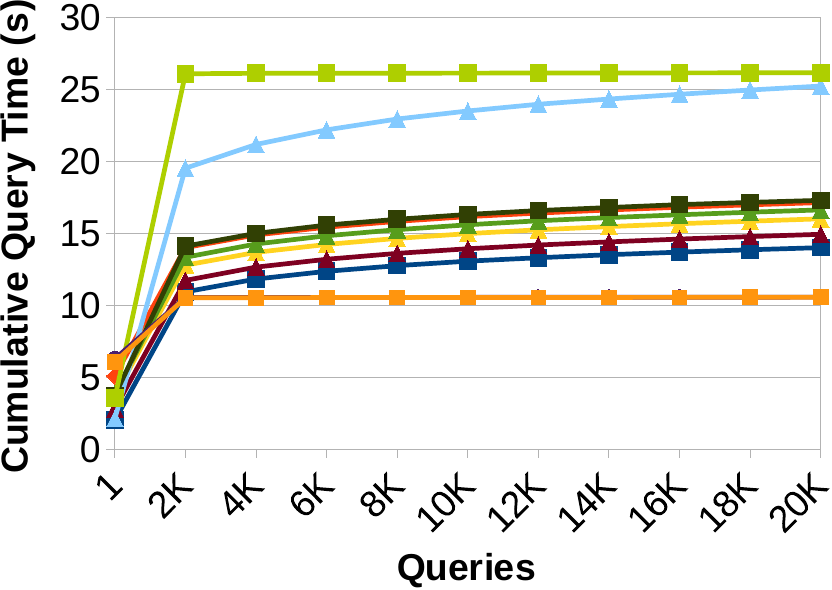}
	    \caption{Zipfian (Read 100\%)}
	    \label{subfig:exp_zip_read_only}
	\end{subfigure}    
 \hspace{10pt}
  \begin{subfigure}[b]{0.3\textwidth}
	   \centering
	   \includegraphics[width=\linewidth]{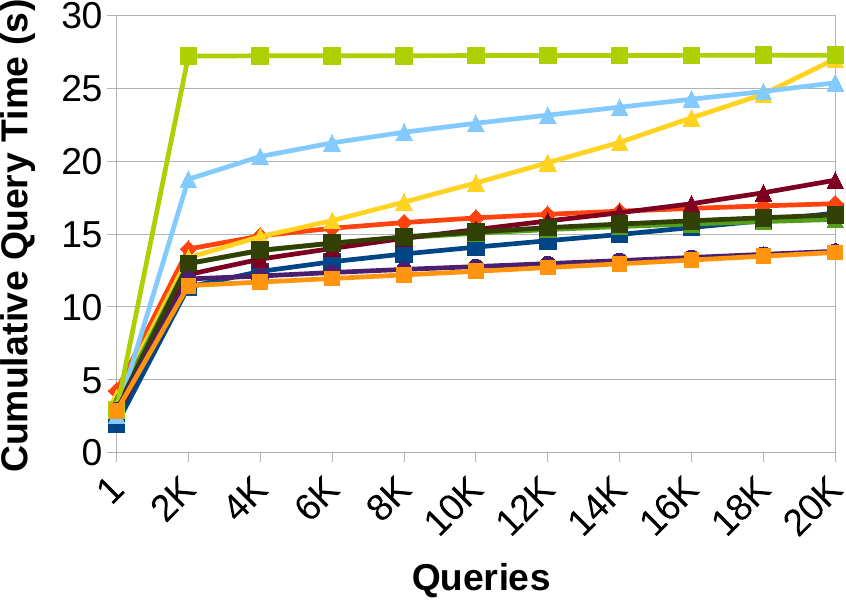}
	    \caption{Zipfian (Read 90\%, Insert 10\%)}
	    \label{subfig:exp_zip_read_90_insert_10}
	\end{subfigure}    
\hspace{10pt}
  \begin{subfigure}[b]{0.3\textwidth}
	   \centering
	   \includegraphics[width=\linewidth]{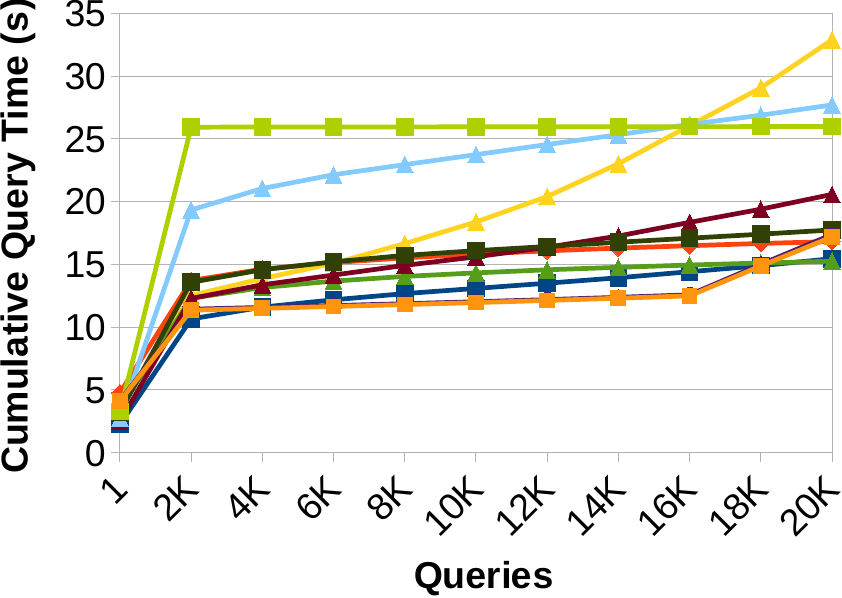}
	    \caption{Zipfian (Read 80\%, Insert 20\%)}
	    \label{subfig:exp_zip_read_80_insert_20}
	\end{subfigure}    
\caption{Evaluation on YCSB workloads }
\label{fig:eval_YCSB}
\end{figure*}

Both \LAI{} and \LAI$_{wofc}$ perform 
better than 
the other adaptive indexes for \textit{Random, Sequential Random, Sequential Alternate, Sequential Inverse} and  \textit{Sequential Overlap} queries (Figs. \ref{subfig:eval_random} -- \ref{subfig:eval_seq_over}). \sdn{So, in these workloads, \LAI$_{wofc}$ provides the best performance even when queries do not get correctly forecasted using our workload prediction approach. The reason for this  is briefly described in Sec \ref{subsubsec:eval_query_workload_pred}.} 


For $ZoomIn$ (Fig. \ref{subfig:eval_zoomin}), the cumulative execution time for \textit{CRACK} was large (around 350 seconds) compared to the other approaches (around 5.5 seconds). Due to this reason, when we plotted them, it was hard to distinguish the performance of the other approaches. Hence, we removed \textit{CRACK} from this figure. Similar to the previous experiments, here also both \LAI{} and \LAI$_{wofc}$ have a high initialization time, and as more queries are submitted its performance nearly converges to that of the other adaptive indexes. From Fig. \ref{subfig:zoomin_query}, we see that after the first query, all the subsequent queries entirely falls within the first query, leading us to \textbf{Case 2}. In this case all the subsequent queries are completely answered from the index built from the first query. Due to this reason the cumulative execution time of both \LAI{} and \LAI$_{wofc}$ remains almost the same as the first query. Hence, there is no advantage of query workload prediction in this scenario as well.

In case of \textit{Sequential ZoomIn} (Fig. \ref{subfig:eval_seq_zoomin}) \LAI{} performs the best, whereas the performance of \LAI$_{wofc}$ is not good enough compared to the other indexes (in spite of having similar execution times until 1K queries), showcasing the advantage that we get from query workload prediction. 
In \LAI, as we were able to forecast the queries in the workload with reasonably high accuracy, this resulted in only 2 queries falling under \textbf{Case 1(ii)} and the remaining under \textbf{Case 2}. Using query workload prediction, the reduction in the number of queries falling under \textbf{Case 1(ii)} in \LAI{} has resulted in  significantly reduced query execution times. \skd{Although around 19.9K queries fall under \textbf{Case 2} for both  \LAI{} and \LAI$_{wofc}$, answering them using \ULI{}
is quite fast and it takes negligible amount of time compared to the cumulative execution time. Hence, the time values appear as 0 in Fig. \ref{subfig:eval_seq_zoomin_case_time}.} 

\subsubsection{\textbf{Evaluating query workload prediction.}}
\label{subsubsec:eval_query_workload_pred}
To show the benefits of query workload prediction, we also measure 
the total execution time of these four workloads  that each of the query spends in different cases (shown in Fig. \ref{fig:eval_synthetic_query_workloads_casewise}). From Figs. \ref{subfig:seq_zoomin_query} -- \ref{subfig:periodic_query}, we see that both the endpoints of the queries from these four workloads follow a pattern and are thus quite predictable. Thus, query workload forecasting offers an advantage in these situations. 
For \textit{Zoomout, Sequential ZoomOut} and \textit{Periodic} workloads (Figs. \ref{subfig:eval_zoomout} -- \ref{subfig:eval_periodic}) we  can 
observe significant improvement in query execution time with \LAI{} compared to \LAI$_{wofc}$. \sdn{Also, from Figs. \ref{subfig:eval_random} to  \ref{subfig:eval_seq_over} we see that the performance of \LAI{} and \LAI$_{wofc}$ are similar. In all these cases, either all or most  of the query endpoints were randomly generated. This makes the future query prediction hard for \LAI$_{wofc}$ in these scenarios. However, the predicted queries can still help by updating the index for arbitrary query ranges, which is later utilized when actual queries are submitted. These results show that the performance of \LAI$_{wofc}$ is no worse that \LAI{} irrespective of predictability of the future queries.}

\subsubsection{\textbf{Evaluating \LAI{} and \LAI$_{wofc}$ on YCSB workloads.}}
The evaluation of \sdn{\LAI{} and} \LAI$_{wofc}$ over YSCB workload is shown in Fig. \ref{fig:eval_YCSB}. Here we execute range and insert queries over 100M integer keys where the query workload are generated from YCSB benchmark. We use three different workloads each for Uniform and Zipfian distribution, as shown in Table \ref{tab:workloads}, by varying the read and insert ratio as: \textbf{Read Only}, \textbf{Read 90\%-Insert 10\%} and \textbf{Read 80\%-Insert 20\%}. For 
each spline point a buffer of size 256 is maintained to accommodate the inserts. \sdn{Since, it is 
\srt{challenging to correctly} predict the values to be inserted in future, in \LAI{} we only use prediction for the future read queries based on past read queries, and skip the \srt{prediction} for insert queries. 
}
We evaluate our approach with the same set of existing approaches and report the cumulative query execution time (both read and insert queries) for 20K queries. \sdn{Here too, we notice that the performance of \LAI{} and \LAI$_{wofc}$ are similar, except for Uniform (Read 80\%-Insert 20\%) as shown in Fig. \ref{subfig:exp_uniform_read_80_insert_20}.} We observe that our approach outperforms all the other approaches for almost all the workloads, except for Zipfian (Read 80\%-Insert 20\%).

In case of Uniform distribution, for \textbf{Read 100\%} queries, learned model gets created until the first 2K queries \sdn{in case of both \LAI{} and \LAI$_{wofc}$}. It takes just over 10 secs to answer the first 2K queries, beyond which the execution time remains constant, suggesting that all queries after 2K queries utilize the indexes built so far for execution. For \textbf{Read 90\%-Insert 10\%}, we see that the cumulative query execution time gradually increases as more and more queries are submitted. The first query takes most time to execute for \sdn{both \LAI{} and } \LAI$_{wofc}$ compared to the other approaches. But their performance improves as more queries are submitted. There is a sudden jump in the cumulative execution time between 18K--20K queries, suggesting that most of the models had to be rebuilt during this phase. The performances of \sdn{\LAI{} and }\LAI$_{wofc}$ for \textbf{Read 80\%-Insert 20\%} is almost similar to the previous scenario except the fact that there is a sudden jump in the cumulative query time between 10K--12K queries \sdn{(for \LAI$_{wofc}$) and between 10K--12K queries (for \LAI{})}. This suggests that most of the learned index model had to be retrained much earlier compared to the previous scenario, due to increased number of inserts. \sdn{Also, we see improvement in execution time when we predict the queries in \LAI{} compared to \LAI$_{wofc}$ for 12K queries.}
The behavior for Zipfian distribution is almost similar to that of Uniform distribution. We skip these results due to space constraint.

\section{Conclusion}
\label{sec:conclusion}
Both \textit{learned indexes} and \textit{adaptive indexes} have their own advantages. \sdnj{We believe that implementing adaptive indexing in big data systems can enhance performance.} We have proposed \LAI{} that incorporate learned indexes for the purpose of adaptive indexing and thus bring the advantages of the both. To support updates and improve index build performance, we have introduced an updatable learned index \ULI. 
\LAI{} incorporates an workload prediction approach that helps improve its cumulative query execution time. We have evaluated \LAI{} against other adaptive indexes against various workloads and show that it outperforms them in most cases. 
\sdn{As a future work, we plan to investigate \srt{how to} improve the performance of \LAI{} on workloads, such as \textit{ZoomOut and Periodic} and support deletion operation. We also aim to extend \LAI{} to handle multi-dimensional queries.}



\bibliographystyle{IEEEtran}

\bibliography{main_bib} 

\end{document}